\documentclass[prl,twocolumn,amsmath,amssymb,superscriptaddress]{revtex4-1}

\pdfoutput=1

\usepackage[pdftex]{graphicx}
\usepackage[pdftex,colorlinks=true,urlcolor=blue,linkcolor=black]{hyperref} 
\usepackage{dcolumn}
\usepackage{bm}
\usepackage{datetime}
\usepackage{float}
\usepackage{color}
\usepackage{units}
\usepackage{amsmath}
\usepackage{xcolor}

\newcommand{\ket}[1]{\vert #1 \rangle}	

\def\EF{{E_\mathrm{F}}}
\def\kF{{k_\mathrm{F}}}
\def\EB{{E_\mathrm{B}}}

\def\dE{{\Delta E}}
\def\dEEB{{\Delta E/E_\mathrm{B}}}
\def\lnkfa{{\ln{(k_\text{F}a_\text{2D})}}}
\def\kfa{{k_\text{F}a_\text{2D}}}
\def\TTF{{T/T_\textrm{F}}}

\def\ket#1{\lvert#1\rangle}

\begin{document}

\title{High temperature pairing in a strongly interacting two-dimensional Fermi gas}

\author{P. A. Murthy}
\email{murthy@physi.uni-heidelberg.de}
\affiliation{Physics Institute, Heidelberg University, Heidelberg, Germany}

\author{M. Neidig}
\affiliation{Physics Institute, Heidelberg University, Heidelberg, Germany}

\author{R. Klemt}
\affiliation{Physics Institute, Heidelberg University, Heidelberg, Germany}

\author{L. Bayha}
\affiliation{Physics Institute, Heidelberg University, Heidelberg, Germany}

\author{I. Boettcher}
\affiliation{Department of Physics, Simon Fraser University, Burnaby, BC, Canada}

\author{T. Enss}
\affiliation{Institute for Theoretical Physics, Heidelberg University, Heidelberg Germany}

\author{M. Holten}
\affiliation{Physics Institute, Heidelberg University, Heidelberg, Germany}

\author{G. Z\"{u}rn}
\affiliation{Physics Institute, Heidelberg University, Heidelberg, Germany}

\author{P. M. Preiss}
\affiliation{Physics Institute, Heidelberg University, Heidelberg, Germany}

\author{S. Jochim}
\affiliation{Physics Institute, Heidelberg University, Heidelberg, Germany}

\date{\today}

\begin{abstract}
We observe many-body pairing in a two-dimensional gas of ultracold fermionic atoms at temperatures far above the critical temperature for superfluidity. For this, we use spatially resolved radio-frequency spectroscopy to measure pairing energies over a wide range of temperatures and interaction strengths. In the strongly interacting regime where the scattering length between fermions is on the same order as the inter-particle spacing, the pairing energy in the normal phase significantly exceeds the intrinsic two-body binding energy of the system and shows a clear dependence on local density. This implies that pairing in this regime is driven by many-body correlations, rather than two-body physics. We find this effect to persist at temperatures close to the Fermi temperature which demonstrates that pairing correlations in strongly interacting two-dimensional fermionic systems are remarkably robust against thermal fluctuations. 
\end{abstract}

\keywords{Suggested keywords}
\maketitle

Fermion pairing is the key ingredient for superconductivity and superfluidity in fermionic systems  \cite{Yang1962}. In a system with s-wave interactions, two fundamentally different scenarios can occur: In the first one, as realized for weakly attractive fermions that are described by the theory of Bardeen-Cooper-Schrieffer (BCS), formation and condensation of pairs both take place at the same critical temperature ($T_\textrm{c}$) \cite{Bardeen1957}. While the mean-field BCS picture successfully describes a large class of superconducting materials, strongly correlated electron systems may follow a different pattern. In this second case, preformed pairs suppress the density of states at the Fermi surface at temperatures exceeding the critical temperature. Finding a description of this so-called 'pseudogap' phase, especially for two-dimensional (2D) systems, is thought to be a promising route to understanding the complex physics of unconventional superconductivity \cite{Chen2009, Mueller2017}. This is of particular interest in the context of recent ARPES experiments on Iron Chalcogenide films \cite{Lee2014, Kasahara2016}, where the combined effect of strong s-wave interactions and reduced dimensionality has been shown to result in preformed pairing and superconductivity at remarkably high temperatures.  

\begin{figure*} [ht!]
	\includegraphics[width=0.65\textwidth]{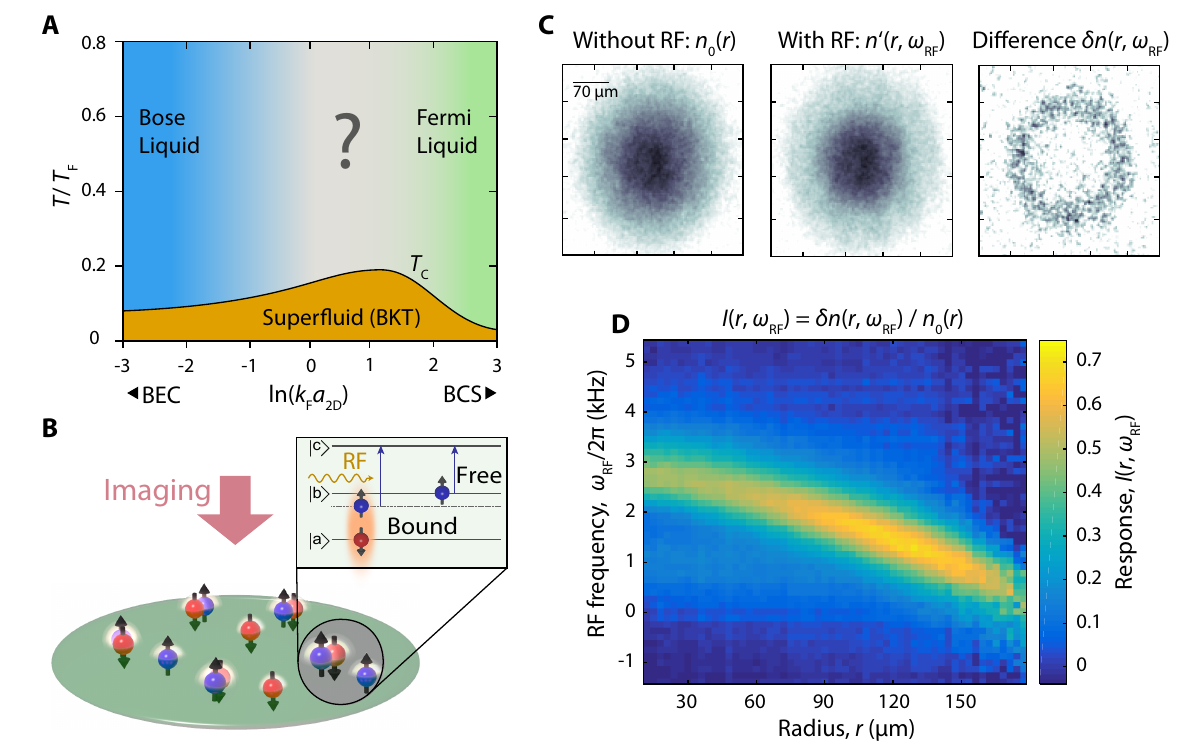}
	\caption{\textbf{Exploring fermion pairing in a strongly interacting 2D Fermi gas.} \textbf{A:} Schematic phase diagram of the BEC – BCS crossover. In this work we investigate the nature of pairing in the normal phase of the crossover regime between the weakly interacting Bose and Fermi liquids. \textbf{B:} Illustration of radio-frequency (RF) spectroscopy of a 2D two-component Fermi gas. Pairing and many-body effects shift the atomic transition frequencies between the hyperfine states $\ket{b}-\ket{c}$, which results in observable signatures in the RF response of the system. \textbf{C:} Absorption images of the cloud without RF (reference) and with RF at a particular frequency, and the difference between the two images. The ring feature in $\delta n(r)$  reveals the density dependence of the RF response. \textbf{D:} Spatially resolved spectral response function reconstructed from absorption images taken at different RF frequencies. At low temperatures in the spin-balanced sample, the occupation of the free particle branch is too low to be observable, which makes it difficult to distinguish between mean-field shifts and pairing effects. }
	\label{fig:method}
\end{figure*}

The Bose--Einstein Condensation (BEC)-BCS crossover of ultracold atoms constitutes a versatile framework to explore the normal phase of strongly correlated fermions (Fig. \ref{fig:method} A). The crossover smoothly connects two distinct regimes of pairing: the BEC regime of tightly bound molecules and the BCS regime of weakly bound Cooper pairs. In 2D (unlike 3D) systems with contact interactions, a two-body bound state with binding energy $\EB$ exists for arbitrarily small attraction between the atoms. The interactions in the many-body system are captured by the dimensionless parameter $\lnkfa$ where $\kF$ is the Fermi momentum and $a_\text{2D}$ is the 2D scattering length. As we tune the interaction strength from the BEC (large negative  $\lnkfa$)  to the BCS side (large positive  $\lnkfa$), the behavior of the system smoothly changes from bosonic to fermionic character \cite{Levinsen2015}. The fascinating strongly interacting region lies in between these two weakly interacting limits where $a_\text{2D}$ is on the same order as the inter-particle spacing ($\sim \kF^{-1} $). In our previous works, we used a matterwave focusing method to measure the pair momentum distribution of a 2D Fermi gas across the crossover and observed the Berezinskii--Kosterlitz--Thouless (BKT) transition to a superfluid phase at low temperatures \cite{Ries2015, Murthy2015}. An outstanding question concerns the nature of the normal phase above the critical temperature - is it a gapless Fermi Liquid of quasiparticles or a gapped liquid of preformed pairs \cite{Torma2016}? While previous cold atom experiments have explored this regime both in 3D \cite{Nascimbene2011, Gaebler2010, Perali2011, Sagi2015, Schunck2007} and 2D \cite{Sommer2012, Feld2011} systems,  a consensus is yet to emerge. 

Here, we address these questions by studying the normal phase of such a 2D ultracold Fermi gas trapped in a harmonic potential. The underlying potential leads to an inhomogeneous density distribution and by making use of a local density approximation, we can directly measure the density dependence of many-body properties. We perform our experiments with a two-component mixture of ${^6}$Li atoms with approximately $3 \times 10^4$ particles per spin state that are loaded into a single layer of an anisotropic harmonic optical trap. The trap frequencies $\omega_z \approx 2\pi \times 6.95\,$kHz and $\omega_z \approx 2\pi \times 22\,$Hz in the axial and radial directions result in an aspect ratio of about 300:1. The kinematic 2D regime is reached by ensuring that the thermodynamic energy scales, temperature ($T$) and chemical potential ($\mu$), are smaller than the axial confinement energy. We tune the scattering length $a_\text{2D}$  by means of a broad magnetic Feshbach resonance \cite{Zurn2013}.

To investigate fermion pairing in our system, we use radio--frequency (RF) spectroscopy. The idea underlying this technique is that the bare atomic transition frequencies between distinct hyperfine states are shifted due to interactions or pairing between atoms in an ensemble (see Fig. \ref{fig:method} B). For this we start with a mixture of atoms in hyperfine states $\ket{a}$  and $\ket{b}$. An RF pulse transfers atoms from state $\ket{b}$ to a third unoccupied hyperfine state $\ket{c}$, and we subsequently image the remaining density distribution in $\ket{b}$. By measuring the spectral response to an applied RF pulse, we gain insight into pairing and correlations in the many-body system. The response functions are affected by interactions between atoms in the initial and final states during the RF pulse. In ${^6}$Li gases, this is due to overlapping Feshbach resonances between the three lowest hyperfine states (labeled $\ket{1}\ket{2}$  and $\ket{1}\ket{3}$). We perform our measurements on two different initial state mixtures ($\ket{a}\ket{b} \equiv \ket{1}\ket{2}$   or $\ket{a}\ket{b} \equiv \ket{1}\ket{3}$), choosing them in such a way that the final state interactions are small for each given interaction strength \cite{SOM2017}.

In our inhomogeneous 2D system, the Fermi energy depends on the local density $n(r)$ in each spin state according to $\EF = \hbar^2 \kF^2 / 2m = (2\pi\hbar^2/m) n(r)$, where $m$ is the mass of a $^6$Li atom \cite{EFnote2017}. As a consequence, the thermodynamic quantities $\TTF$ and $\lnkfa$ also vary spatially across the cloud. We apply the thermometry developed in our previous work \cite{SOM2017, Boettcher2016}  to extract these local observables. We measure the local spectral response \cite{Shin2007} by choosing a RF pulse duration ($\tau_\textrm{RF} = 4\,$ms) that is sufficiently short to prevent significant diffusion of transferred atoms, but also sufficiently long that we obtain an adequate Fourier limited frequency resolution $\delta\omega_\textrm{RF} \approx 2\pi \times 220\,$Hz (see \cite{SOM2017} for details). In Fig. \ref{fig:method} C, we show a typical absorption image of the 2D cloud which is used as a reference and one with a RF pulse applied at a particular frequency. The difference between the two images features a spatial ring structure, which qualitatively shows that for a given frequency, the depletion of atoms in initial state $\ket{b}$ occurs at a well-defined density/radius. By performing this measurement for a range of RF frequencies, we can tomographically reconstruct the spatially resolved spectral response function
\begin{equation}
I(r,\omega_\textrm{RF}) = (n_0(r) - n'(r,\omega_\textrm{RF}))/n_0(r),
\end{equation}
where $n_0(r)$  and $n'(r,\omega_\textrm{RF})$  are the density distribution of atoms in state $\ket{b}$ without and with the RF pulse. An example of the tomographically reconstructed spectra is shown in Fig. \ref{fig:method} D. We see that the frequency of maximum response depends smoothly on the radius and thereby the local density. These density-dependent shifts may arise either from pairing in the system or from interaction induced mean-field shifts. One way to distinguish between the two contributions is to measure the RF transitions from both paired and unpaired atoms to the free particle continuum \cite{SOM2017}. However, we find that in the temperature regime ($\TTF < 1.5$) explored in our experiments, the thermal occupation of the free branch is too small to observe.

\begin{figure} [ht!]
	\includegraphics {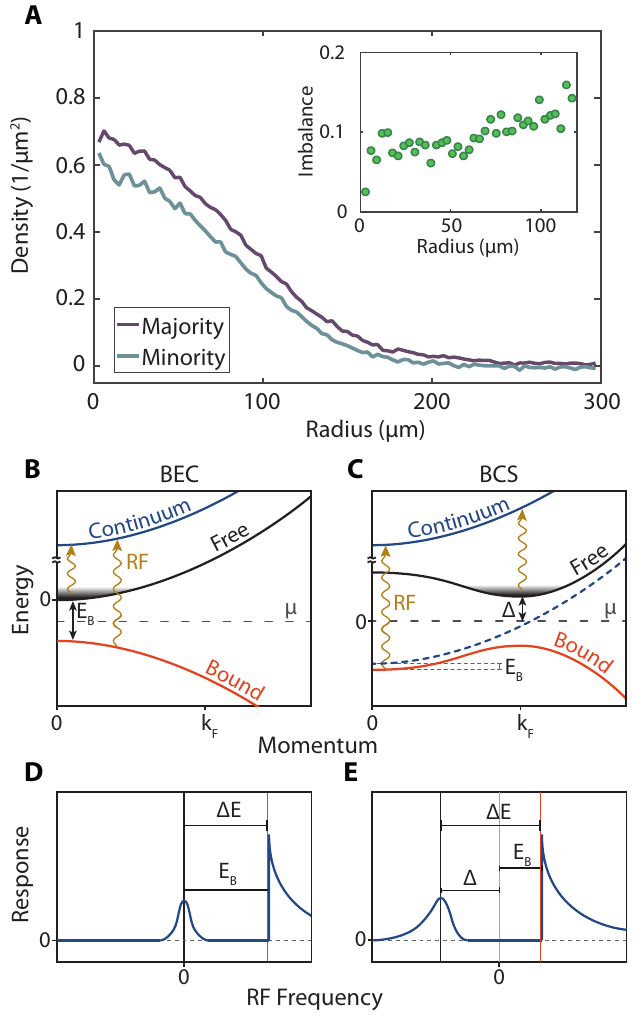}
	\caption{\textbf{Quasiparticle spectroscopy in the BEC and BCS limits.} \textbf{A:} We create a slightly imbalanced mixture of hyperfine states to artificially populate the free particle branch. The density distributions of the majority and minority spins are shown, and the corresponding local imbalance (inset). \textbf{B, C:} Schematic illustration of single particle dispersion relations in the BEC and BCS limits at zero temperature. Paired atoms reside in the lowest branch (bound), and are transferred to the continuum of unoccupied states. The excess majority atoms are unpaired and occupy the upper quasiparticle (free) branch in the spectrum at $k \sim 0$ (BEC) and $k \sim \kF$ (BCS).  \textbf{D, E:} The transition of paired atoms into the continuum yields an asymmetric response with a sharp threshold in the RF spectral function. The quasi-particle transition contributes another peak. Their relative difference yields the pairing energy $\dE$ which reveals the distinction between two-body ($\dE \sim \EB$) and many-body pairing ($\dE > \EB$) in the two limits.}
	\label{fig:imbalance}
\end{figure}
In order to achieve a sufficient population of the unpaired branch, we apply the quasiparticle spectroscopy method pioneered in \cite{Schirotzek2008} for the measurement of the superfluid gap of a 3D Fermi gas. Although our system is in the normal phase, the same technique can be used to determine the pairing gap. The key idea of this method lies in creating a slightly spin-imbalanced mixture so that the excess majority atoms necessarily remain unpaired due to the density mismatch. These unpaired atoms (or dressed quasiparticles) contribute a second absorption maximum in the RF response function besides the one from pairs. Since mean-field interactions shift the whole spectrum \cite{SOM2017}, the difference $\dE$ between the two branches is manifestly independent of it and corresponds only to the pairing energy of the system. In our experiments, we create a slight spin-imbalance $P=(n_{\ket{b}} -  n_{\ket{a}} )/(n_{\ket{b}} + n_{\ket{a}}) \lesssim 0.15$ using a sequence of Landau-Zener sweeps, where $n_{\ket{b}}$⟩  and $n_{\ket{a}}$  are densities in hyperfine states $\ket{b}$ and $\ket{a}$. We show typical density profiles of majority and minority components in Fig. \ref{fig:imbalance} A.

\begin{figure*} [ht!]
	\includegraphics {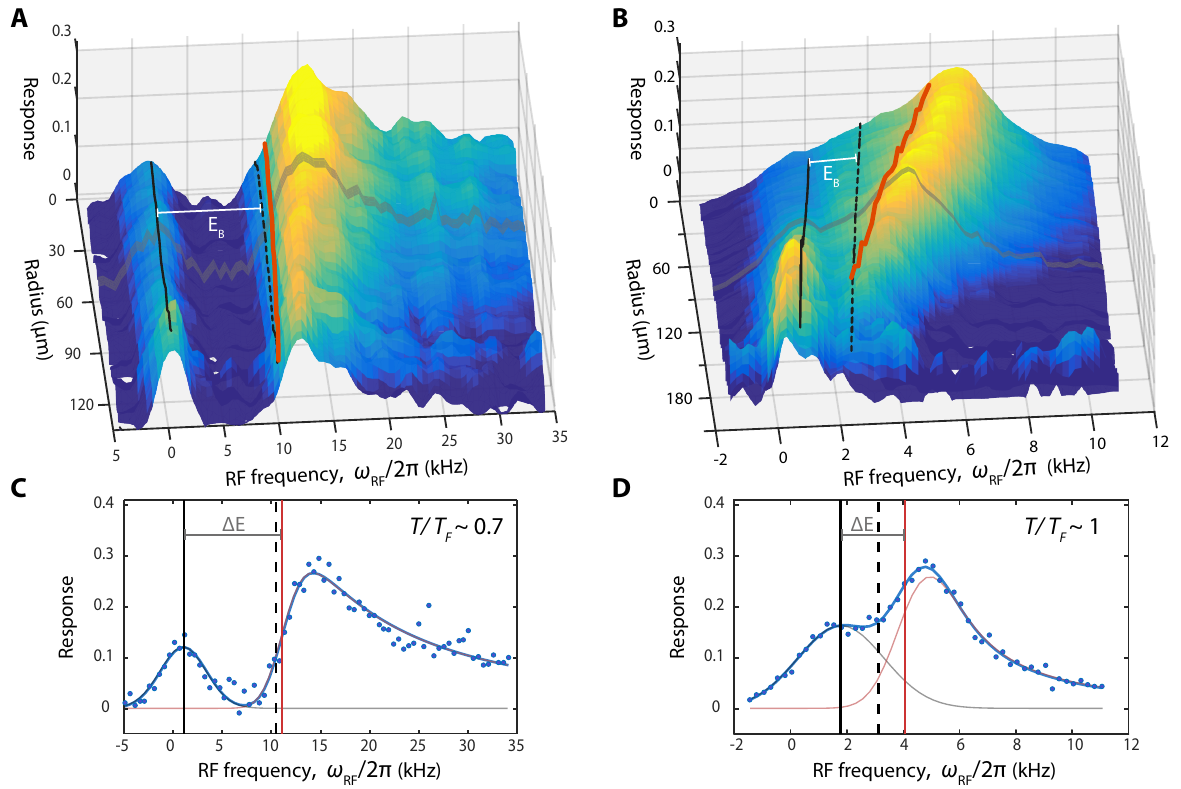}
	\caption{\textbf{From two-body dimers to many-body pairing.} The spatially resolved response function $I(r,\omega_\textrm{RF})$ shows qualitatively different behavior for two different scattering lengths. \textbf{A, B:} $I(r,\omega_\textrm{RF})$ for central $\lnkfa \sim -0.5$ and $1.0$, respectively. The 3D visualization is obtained using a linear interpolation between 3000 data points each of which is an average of 30 realizations. The black solid line is the peak position of the free branch, the orange line is the threshold position of the bound branch and the black dashed line is displaced from the free peak by the two-body binding energy $\EB$.  The energy difference between free and bound branches is the pairing energy $\dE$, which is seen to agree with $\EB$ in \textbf{A} (BEC regime), but signigicantly exceeds $\EB$ in \textbf{B} (crossover regime). In addition,$\dE$ is strongly density dependent in \textbf{B}, implying that it originates from many-body correlations. \textbf{C, D:} Local spectra at a fixed radius corresponding to a homogeneous system with $\TTF \sim 0.7$ and $1$ respectively (gray region). The solid blue curves are the fits to the data; the black and red curves are gaussian and threshold fits to the two branches. }
	\label{fig:MBP}
\end{figure*}

The pairing energy $\dE$  between the two branches distinguishes between two different pairing scenarios. If $\dE$ coincides with the two-body bound state $\EB$, we are in the two-body regime. In contrast, we associate the situation of a density ($\EF$)-dependent $\dE$ exceeding $\EB$ with many-body pairing. In Fig. \ref{fig:imbalance} B and C, we illustrate these two scenarios using idealized single-particle dispersion relations in the BEC and BCS limits at zero temperature. We provide a brief theoretical account of pairing in these two limits in \cite{SOM2017}. The crucial difference between the two cases lies in the occupation of the quasiparticle branch which occurs preferentially at $k \sim 0$ in the BEC \cite{Barth2014} and at $k \sim \kF$ in BCS \cite{Fischer2014} regimes. While the corresponding RF spectra, shown in Fig. \ref{fig:imbalance} D and E, appear qualitatively similar, the value of $\dE$ reveals the fundamental difference in the nature of pairing in the two regimes. We note that the actual dispersion relations at strong interactions and high temperatures – which determine the RF response in the experiment – do not necessarily follow this mean-field description \cite{Bauer2014,Vitali2017}. However, the general criteria to distinguish two-body and many-body pairing scenarios using $\dE$ remain valid.

In Fig. \ref{fig:MBP} A and B, we show the measured spectra $I(r,\omega_\textrm{RF})$ for magnetic fields $670\,$G and $690\,$G using a $\ket{1}\ket{3}$ mixture, which corresponds to central values of $\lnkfa \sim -0.5$ and $\lnkfa \sim 1$, respectively. The response from unpaired quasiparticles appears at frequency $\omega_\textrm{RF} \sim 0$, while the pairing branch with an asymmetric lineshape appears at larger frequencies. Examples of spectra at fixed radii are shown in Fig. \ref{fig:MBP} C and D. We fit these local spectra with a combined fit function that includes a symmetric Gaussian (for the quasiparticle peak) and an asymmetric threshold function (for the paired peak) that is convolved with a Gaussian to account for spectral broadening arising from finite RF frequency resolution and final state effects \cite{Langmack2012}. The choice of fit function has a systematic effect on the quantitative results presented here, which cannot be eliminated at this point since a reliable theoretical prediction of the shape of the spectral function only exists in the weakly coupled BEC \cite{Barth2014} and BCS \cite{Fischer2014} limits. We present a detailed account of our data analysis in \cite{SOM2017}. 

\begin{figure} [ht!]
	\includegraphics {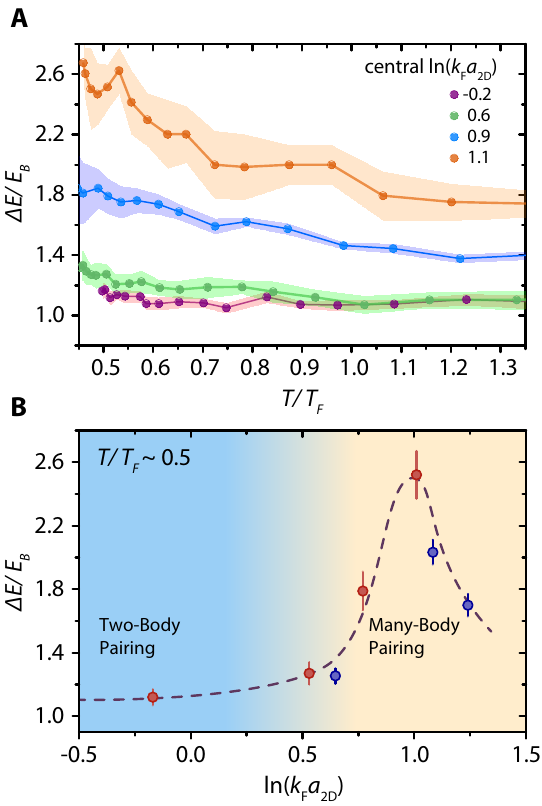}
	\caption{\textbf{Normal phase in the 2D BEC-BCS crossover regime.} \textbf{A:} Pairing energy $\dE$ in units of $\EB$ plotted as a function of $\TTF$ for different interaction strengths (central $\lnkfa$) . \textbf{B:} Many-body induced high temperature pairing. We plot $\dEEB$ as a function of $\lnkfa$ for fixed ratio $\TTF \sim 0.5$. Red and blue circles correspond to measurements taken with $\ket{1}\ket{3}$ and $\ket{1}\ket{2}$  mixtures. The dashed black line is a guide to the eye. The errors indicated as shaded bands in \textbf{A} and bars in \textbf{B} are obtained from the fitting procedure explained in \cite{SOM2017}. For $\lnkfa \leq 0.5$ (strong attraction) we have $\dEEB \sim 1$ with negligible density-dependence, indicating two-body pairing. For larger $\lnkfa$ (less attraction), $\dEEB$ significantly exceeds 1 and reaches a maximum of 2.6 before showing a downward trend. At $\lnkfa \sim 1$, we have a critical temperature of $T_\textrm{c}  \approx 0.17\,T_\textrm{F}$ \cite{Ries2015}, which indicates the onset of many-body pairing at temperatures several times $T_\textrm{c}$. }
	\label{fig:crossover}
\end{figure}

At a qualitative level, the main observations from Fig. \ref{fig:MBP} are the following. Both branches in the spectra show density dependence, which can partially be attributed to a mean-field shift. Adding the binding energy $\EB$ to the quasi-particle branch yields the two-body expectation for the threshold position. This picture is applicable to the whole spectrum in Fig \ref{fig:MBP} A, which corresponds to a measurement on the BEC side of the crossover. In contrast, for the spectrum displayed in Fig \ref{fig:MBP} B, corresponding to the crossover regime, we observe $\dE \sim \EB$ only in the outer regions of the cloud where the density is low enough that only the two-body bound state plays a role. Towards the center of the cloud, $\dE$ begins to significantly exceed $\EB$ and shows a strong dependence on the local density ($\EF$), indicating that pairing in this regime is a many-body phenomenon. At very low temperatures, the measurement of $\dE$ is difficult since the occupation of the free branch is too low, as seen in Fig. \ref{fig:method} D. However, we qualitatively observe that the threshold position of the bound branch increases continuously with decreasing temperature, even as we cross the superfluid transition. This indicates that in the crossover regime, a many-body gap opens up far in the normal phase rather than at $T_\text{c}$ as expected from BCS theory. This observation is the first main result of this work.

To quantitatively study the change in the nature of pairing from the BEC to the BCS side, we measure the spectra at different magnetic fields and extract $\dE$ in units of the two-body binding energy $\EB$. In Fig. \ref{fig:crossover} A, we plot the temperature dependence of $\dEEB$ for different interaction strengths, and Fig. \ref{fig:crossover} B shows the variation of $\dEEB$ as a function of $\lnkfa$ for a fixed ratio $\TTF \approx 0.5$. This constitutes an extremely high temperature regime even in the context of ultracold fermion superfluidity, where the largest observed critical temperatures are $T_\textrm{c} / T_\textrm{F} \approx 0.17$ \cite{Ries2015,Murthy2015}. We perform our measurements with both $\ket{1}\ket{2}$ and $\ket{1}\ket{3}$ mixtures (blue and red points in Fig. \ref{fig:crossover} B) in an overlapping interaction regime. The two mixtures differ vastly in their final state interaction strengths and the fact that we observe consistent behavior with both mixtures demonstrates the robustness of the quantity $\dE$ against these final state effects. Details of the experimental parameters used for the two mixtures are tabulated in \cite{SOM2017}.

In Fig. \ref{fig:crossover}, we observe that for $\lnkfa \lesssim 0.5$ the spectra are well-described by two-body physics. In contrast, the pronounced density-dependent gap significantly exceeding $\EB$ for $\lnkfa \geq 0.5$ signals the crossover to a many-body pairing regime. In particular, we observe that $\dEEB$ peaks at $\lnkfa \sim 1$, where $\dE \approx 2.6\EB$  and is a significant fraction of $\EF$ ($0.6\EF$). The identification of this strongly correlated many-body pairing regime and the observation of many-body induced pairing at temperatures several times the critical temperature is the second main result of this work. For larger $\lnkfa$, we see a downward trend in $\dEEB$, and for $\lnkfa \geq 1.5$, we observe only a single branch in the spectra near $\omega_\textrm{RF} \sim 0$, suggesting the absence of a gap larger than the scale of our experimental resolution \cite{SOM2017}. Our qualitative observation of a vanishing gap for weaker attraction is consistent with the picture of the normal phase in the BCS limit being a gapless Fermi liquid \cite{Frohlich2012}.

We now turn to a discussion of our results in the context of current theoretical understanding and previous experimental work. In \cite{Sommer2012}, Sommer et al. performed trap-averaged RF spectroscopy in the 3D-2D crossover and found good agreement with mean-field BCS theory in the regime $\lnkfa \leq 0.5$. In \cite{Feld2011}, Feld et al. observed pairing in the normal phase using momentum-resolved (but trap-averaged) spectroscopy also mostly in the regime $\lnkfa \leq 0.5$. Beyond this previously explored interaction regime where pairing is purely described by two-body physics \cite{Ngampruetikorn2013, Barth2014}, our measurements reveal that many-body effects enhance the pairing energy far above the critical temperature.  In fact, we show that the most strongly correlated region where many-body pairing occurs in the normal phase lies where $\lnkfa$ is close to unity rather than zero. With regard to the long-standing question concerning the nature of the normal phase of a strongly interacting Fermi gas \cite{Torma2016}, our experiments reveal the existence of a large region in the phase diagram where the behavior deviates from both Bose Liquid and Fermi liquid descriptions. Finding a complete description of this strongly correlated phase is an exciting challenge for both theory and experiment. \\
\\

\noindent \textit{Acknowledgements:} We gratefully acknowledge insightful discussions with Meera Parish, Jesper Levinsen, Nicolò Defenu and Wilhelm Zwerger. We would like to thank Thomas Lompe for discussions and for a critical reading of the manuscript. This work has been supported by the ERC consolidator grant 725636, the Heidelberg Center for Quantum Dynamics and is part of the DFG Collaborative Research Centre “SFB 1225 (ISOQUANT)”. IB acknowledges support from DFG und grant no. BO 4640/1-1. P.M.P. acknowledges funding from European Union’s Horizon 2020 programme under the Marie Sklodowska-Curie grant agreement No. 706487.\\

\noindent \textit{Author contributions:} P.A.M and G.Z. initiated the project. P.A.M., M.N., R.K. and M.H. performed the measurements and analysed the data. I.B. and T.E. provided theory support and assistance with preparing the manuscript. P.M.P. and S.J. supervised the project. All authors contributed to the interpretation and discussion of the experimental results.\\ 
\noindent
P.A.M, M.N. and R.K. contributed equally to this work. 
\clearpage


%

\cleardoublepage
\section*{\large Supplementary Materials}
\setcounter{figure}{0}
\renewcommand{\figurename}{Fig.\,S}
\renewcommand{\tablename}{Tab.\,S}


\section{Experimental Design} 
\subsection{Preparing the Sample}
We use the three lowest Zeeman sublevels of the hyperfine ground state of $^6$Li - labeled $\ket{1},\ket{2}$ and $\ket{3}$, as shown in Fig.\,S \ref{fig:S2_RFSpectroscopy} A and B. The starting point for our experiments is an almost pure molecular Bose-Einstein condensate (mBEC) of  $\sim 5\times10^4$ $\ket{1}\ket{2}$ or $\ket{1}\ket{3}$ dimers produced by optical evaporation at magnetic offset fields of $B=\unit[690]{G}$ or $B=\unit[795]{G}$ respectively. The sample is then transferred into an optical standing wave trap (SWT)  (\textit{8}), where we perform additional evaporation to reach the quasi-2D regime with $\sim 3\times 10^4$ atoms in each spin state. At $B=\unit[795]{G}$, the trapping frequencies for the central layers of the SWT are $\omega_\mathrm{r} \approx 2\pi \times \unit[22]{Hz}$ and $\omega_\mathrm{z} \approx 2\pi \times \unit[6.95]{kHz}$. This results in an aspect ratio of $\omega_\mathrm{r}:\omega_\mathrm{z}\approx 1:300$. Here the system is in the kinematic quasi-2D regime, which we experimentally verified in \cite{Ries2015}.

\subsection{Creating a Spin-Imbalanced Mixture}

After evaporation we introduce imbalance by a sequence of Landau-Zener passages at $B = \unit[1000]{G}$ where the interaction strength is relatively weak (see Fig.\,S \ref{fig:S2_RFSpectroscopy} C). In the case of the $\ket{1}\ket{3}$-mixture, we transfer a small fraction of atoms from state $\ket{1}$ to state $\ket{2}$. The ensuing three-body collisions lead to an imbalance between state $\ket{1}$ and state $\ket{3}$ with a majority of the atoms in state $\ket{3}$. For the $\ket{1}\ket{2}$-mixture, we transfer atoms from state $\ket{2}$ to state $\ket{3}$ and invert the resulting imbalance with an additional Landau-Zener passage to end up with the majority in state $\ket{2}$.\\ 
Radial profiles for both the majority and minority component are shown in the main text in Fig. 2 for a $\ket{1}\ket{3}$-mixture. We work with a local polarization $p_\mathrm{loc} = \frac{n_{\ket{3}}-n_{\ket{1}}}{n_{\ket{3}}+n_{\ket{1}}}\lesssim \unit[15]{\%}$. As we operate in the normal phase, we do not observe a phase separation into a balanced core and a polarized wing as one observes in a superfluid \cite{Schirotzek2008,Mitra2016} but rather see a local polarization that varies only weakly over the sample. The heat introduced due to the three-body losses in this imbalance scheme limits the achievable temperatures to about $T/T_\mathrm{F}\gtrsim 0.4$.

\subsection{Radio-Frequency Spectroscopy}
\begin{figure*}[t!]
	\centering
	\includegraphics{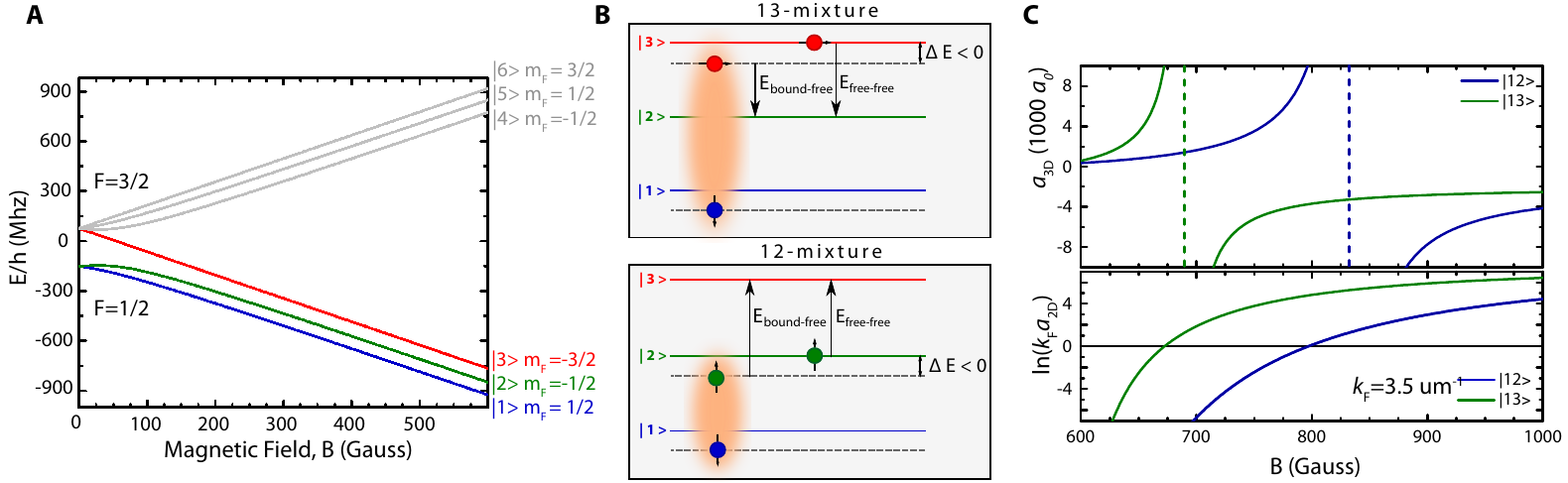}
	\caption{\textbf{A:} Zeeman sublevels of the $^6\mathrm{Li}$ $2^2S_{1/2}$ ground state. Indicated in color are the sublevels experimentally employed. \textbf{B:} Schematic illustration of RF spectroscopy in $^6$Li for both the $\ket{1}\ket{3}$-mixture in the upper panel and the $\ket{1}\ket{2}$-mixture in the lower panel. \textbf{C:} Upper panel: The 3D scattering length $a_\mathrm{3D}$ in units of the Bohr radius $a_0$ is plotted versus magnetic field for the $\ket{1}\ket{2}$-(blue) and $\ket{1}\ket{3}$-mixture (green). Lower panel: The corresponding 2D interaction strength $\ln\left(k_\mathrm{F}a_\mathrm{2D}\right)$ is plotted as a function of magnetic field for a typical Fermi momentum $k_\mathrm{F}=\unit[3.5]{\mu m^{-1}}$.}
	\label{fig:S2_RFSpectroscopy}
\end{figure*}

Fig.\,S \ref{fig:S2_RFSpectroscopy} B schematically shows the RF transitions accessed in our experiments. In the case of a $\ket{1}\ket{3}$-mixture as depicted in the upper panel, we drive the transition from state $\ket{3}$ to $\ket{2}$. Without interactions, the resonance energy for the transition of a free atom is given by $E_\mathrm{free-free}$ which is measured in a spin-polarized, non-interacting sample. If the atom in state $\ket{3}$ is paired with an atom in state $\ket{2}$, the energy level is shifted by the binding energy resulting in a reduced resonance energy $E_\mathrm{bound-free}$. Thus $E_\mathrm{bound-free} < E_\mathrm{free-free}$ and the paired branch is at negative frequencies relative to $E_\mathrm{free-free}$. For the $\ket{1}\ket{2}$-mixture as depicted in the lower panel, the situation is similar. However, here $E_\mathrm{bound-free} > E_\mathrm{free-free}$ and hence the paired branch is at positive RF frequency offsets. For the sake of brevity, we will attribute pairing with positive frequency shifts independent of the mixture throughout this work.  \\
\\
\noindent
\textbf{Final state effects:} 
In an interacting system, the non-interacting free-free transition is not a suitable reference anymore, as interactions - both in the initial and final state - lead to density-dependent mean-field shifts. We address this question, as explained in the main text, by considering only energy differences. In addition, the final state interactions lead to a broadening of the RF transitions which reduces the resolution. This effect can be significant for the $\ket{1}\ket{3}$-mixture where the final state is most likely a short-lived repulsive polaron \cite{Koschorreck2012}. Therefore we make use of the $\ket{1}\ket{3}$- as well as the $\ket{1}\ket{2}$-mixture to minimize the effect of final state interactions. A plot of the broad Feshbach resonances which have considerable overlap between the different mixtures can be seen in Fig.\,S \ref{fig:S2_RFSpectroscopy} C, together with typical values for $\ln\left(k_\mathrm{F}a_\mathrm{2D}\right)$. 
We use the $\ket{1}\ket{3}$-mixture for magnetic fields below the $\ket{1}\ket{3}$-resonance where the final state interaction strength has values of $\ln\left(k_\mathrm{F}a_\mathrm{2D}\right)<-7$, and the $\ket{1}\ket{2}$ mixture for larger fields where the final state  $\ln\left(k_\mathrm{F}a_\mathrm{2D}\right)>4.5$.
The pairing energies measured with both mixtures in the crossover regime show consistent behaviour independent of the mixture used as shown in Fig. 4 B in the main text.\\

\begin{figure}[b!]
	\centering
	\includegraphics{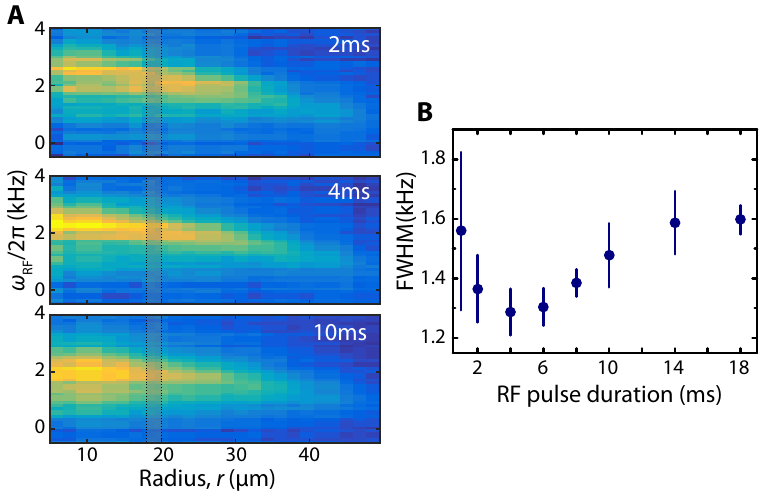}
	\caption{Influence of the RF pulse duration on the measured response spectra at $\unit[854]{G}$. In \textbf{A} the spectra are shown for pulse durations of $\unit[2]{ms}$, $\unit[4]{ms}$ and $\unit[10]{ms}$. Along the cuts marked as the shaded area in the spectra, the FWHM is obtained from fits and shown in \textbf{B}. A pulse duration of $\tau_\mathrm{RF}=\unit[4]{ms}$ gives the optimum trade-off between frequency resolution and the effect of diffusion in the sample.}
	\label{fig:S3_RFPulsedur}
\end{figure}

\noindent
\textbf{Spatially resolved spectroscopy:} To avoid diffusion of the transferred density, we have to use RF pulses which are short compared to the trapping period and image the sample directly after the pulse application. However, short RF pulses lead to a Fourier limited frequency resolution and therefore we have to find the optimum trade-off. To investigate this experimentally, we prepare a $\ket{1}\ket{2}$-mixture at $B = \unit[854]{G}$ and record the RF spectrum for different pulse durations $\tau_\mathrm{RF}$ as can be seen exemplary in Fig. S \ref{fig:S3_RFPulsedur} A. We apply rectangular RF pulses which lead to a Fourier limited width $\Delta \nu \geq 0.886/\Delta t$. For all applied RF pulse durations, the density dependence of the bound-free transition is readily visible. For each pulse duration we adjust the RF power such that a similar fraction of atoms is transferred. This leads to Rabi frequencies between $\unit[60]{Hz}$ for the longest and $\unit[240]{Hz}$ for the shortest pulse durations and hence the effect of power broadening is small compared to the Fourier limit. Binning the spectra over two pixels and taking a cut at a fixed radius, we can then compare the measured widths of the bound-free transition. This is shown in Fig.\,S \ref{fig:S3_RFPulsedur} B for the FWHM of the peak. For short RF pulses on the order of $\unit[1]{ms}$, the width increases as the Fourier limit there is on the order of $\unit[1]{kHz}$. Although the Fourier limit decreases for longer pulse durations, the FWHM increases which we attribute to diffusion of particles during the application of the RF pulse. At around $\tau_\mathrm{RF}\approx\unit[4]{ms}$ the minimal FWHM occurs, which is the pulse length we used throughout this paper. This results in a Fourier limited RF resolution of $\Delta \nu = \unit[222]{Hz}$ with a typical Rabi frequency on the order of $\Omega=\unit[150]{Hz}$.

\section{Data Analysis}

\subsection{Fitting Procedure}

\begin{figure*}[t!]
	\centering
	\includegraphics[width=0.62\textwidth]{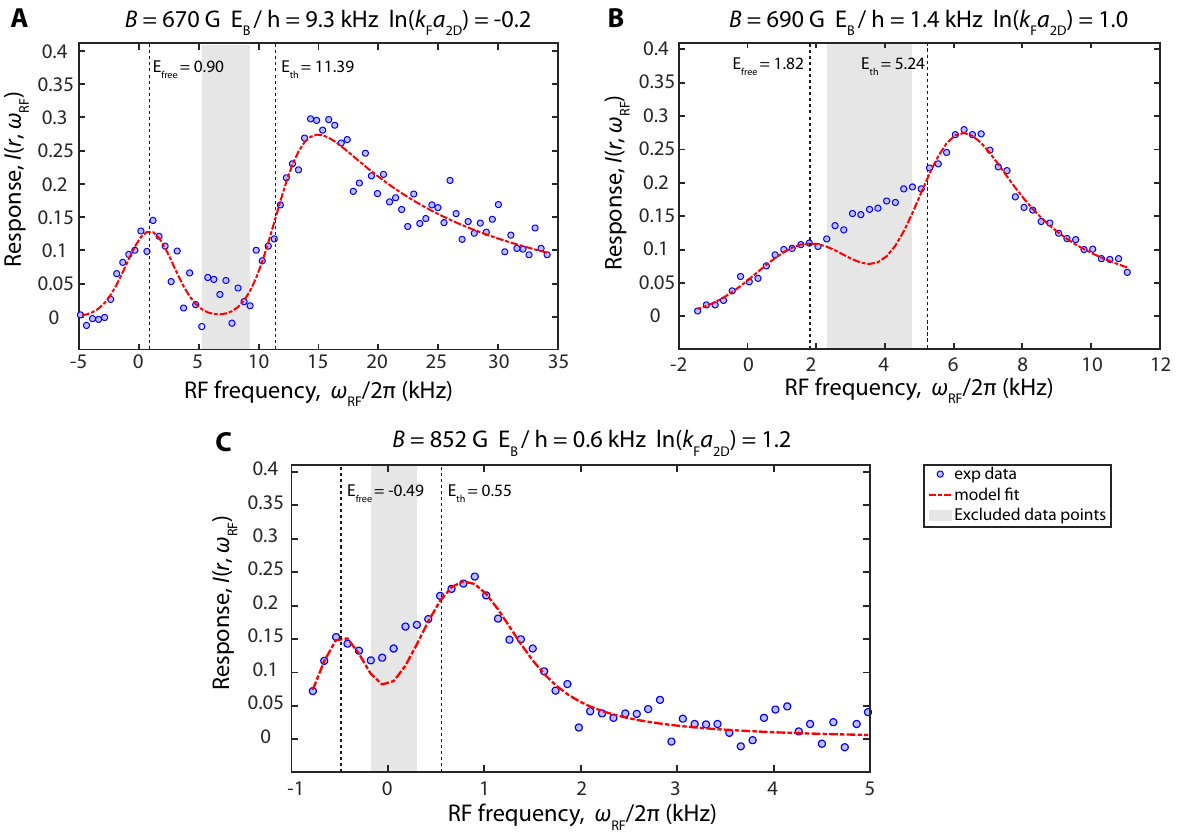}
	\caption{Example cuts through the RF response map at magnetic fields of 672G in \textbf{A}, 692G in \textbf{B} (both in the $\ket{1}\ket{3}$ mixture) and 854G in \textbf{C} (in the $\ket{1}\ket{2}$ mixture). The cuts are chosen such that the local temperature is $T/T_F \approx 0.5$ (c.f. Fig. 4 B of the main text). Shown are the experimental data points (blue circles), a fit of the model function (red dashed line) and the frequency region excluded in the fit (gray area). For the features of interest we see good agreement with the model function at all fields.}
	\label{fig:S5_Fit}
\end{figure*}

We model the RF response at a fixed density and interaction strength by a combination of a symmetric Gaussian profile accounting for the quasi-particle peak and an asymmetric threshold function describing the bound-free peak. For the case of dimers as well as zero temperature BCS theory, the line shape $\Gamma$ of the bound-free peak can be calculated from Fermi\textsc{\char13}s golden rule and is given in terms of the RF frequency $\omega$ by (see Section 3.1) 
\begin{equation} 
\Gamma(\omega) \propto \frac{\theta(\hbar \omega - E_{th})}{\omega^2},
\end{equation}
where $E_{th}=E_B$ denotes the threshold energy for pair breaking, i.e. the binding energy. Due to the $\frac{1}{\omega^2}$ factor, the formula crucially depends on the exact knowledge of the reference frequency (the free-free transition which is set to zero here). This frequency can be affected by mean field shifts, however, which we account for by including a free parameter $\eta$ and replacing $\frac{1}{\omega^2} \rightarrow \frac{1}{(\omega-\eta)^2}$. It is important to note that the maximum of the experimentally recorded RF response does not necessarily coincide with $E_{th}$. The peak position is shifted towards higher energies due to symmetric broadening of the asymmetric response function, as given in particular by the Fourier broadening, as well as final state effects. Similar to what has been presented in \cite{Sommer2012, Langmack2012}, we account for these effects by including an additional factor $\frac{\ln^2(E_{th}/E^f_B)}{\ln^2((\hbar\omega-E_{th})/E^f_B)+\pi^2}$ in terms of the final state two-body binding energy $E^f_B$. This also accounts for the 2D nature of the scattering process. In addition, we convolve the resulting lineshape with a Gaussian profile to account for the aforementioned broadening effects. Thus, our fitting function for the bound-free peak is given by

\begin{align}
\Gamma(\omega) = {\underbrace{\vphantom{\frac{1}{\sqrt{2 \pi\sigma^2_\mathrm{broad}}}}\left( A_{bound}\frac{\theta(\hbar \omega - E_{th})}{(\omega-\eta)^2}\frac{\ln^2(E_{th}/E^f_B)}{\ln^2((\hbar\omega-E_{th})/E^f_B)+\pi^2} \right)}_\text{Threshold function}} \nonumber \\
* {\underbrace{\left( \frac{1}{\sqrt{2 \pi\sigma^2_\mathrm{broad}}}e^{-\frac{\omega^2}{2\sigma_\mathrm{broad}^2}}\right)}_\text{Gaussian broadening}}
\end{align}

where the threshold energy $E_{th}$ relative to the non-interacting free-free transition energy, the amplitude of the bound peak $A_{bound}$, the frequency offset $\eta$ and the width $\sigma_{broad}$ of the Gaussian broadening are kept as free parameters, while the final state binding energy $E^f_B$ is fixed for each magnetic field.\\

\noindent
We observe the broadening to be mostly given by the Fourier limit when using the $\ket{1}\ket{2}$-mixture. However, there is additional broadening in the $\ket{1}\ket{3}$-mixture, possibly due to a short lived repulsive polaron final state \cite{Koschorreck2012}. The model describes the RF response reasonably well in the full crossover, as presented in Fig.\,S \ref{fig:S5_Fit}, and gives a consistent estimate for the onset of pairing throughout the crossover. However, for an exact quantitative determination of the paring gap, a model capturing the spectra of the many-body system is needed, but is currently not available.\\
The observed quasi-particle peak is mostly symmetric and we model it by a Gaussian profile with the peak position, the amplitude and the width as free parameters. We interpret the peak position as the quasiparticle energy. In addition to the signal stemming from bound particles and quasi-particle excitations in the central layer of our 2D trap, we observe additional features at intermediate frequencies which we attribute partly to a small fraction of the atoms being in adjacent layers of the SWT. Therefore, we exclude the corresponding frequency range from the fits.

\subsection{Thermometry}
We extract the temperature of the system with the method established in \cite{Boettcher2016} by fitting reference equations of state (EOS) to the density distribution. Here we slightly modify the thermometry scheme by including an effective mass interpolation in the Boltzmann EOS: 
\begin{equation}\label{eqn:boltz}
n_{ 0 }= \frac{ \alpha }{ \lambda_T^2 } e^{ \alpha\beta\mu},
\end{equation}
where $\lambda_T$ is the thermal de Broglie wavelength and $\beta=\frac{1}{k_\mathrm{B}T}$.
Depending on the magnetic field $B$, the effective temperature exponent $\alpha$ specifies whether the particles are treated as dimers or as free atoms. For $\ln\left(k_\mathrm{F}a_\mathrm{2D}\right) \leq 0$ we assume the system to be molecular and use $\alpha=2$ whereas for $\ln\left(k_\mathrm{F}a_\mathrm{2D}\right) \geq 2$ the system is fermionic with $\alpha=1$. Between these two limits we interpolate $\alpha$ linearly as described in \cite{Ries2015}. Typical statistical errors on the extracted temperatures are on the order of \unit[1-2]{\%}.\\

\noindent
\textbf{Local density approximation:} Due to a spatially inhomogeneous density distribution the thermodynamic quantities $T/T_\mathrm{F}$ and $\ln(k_\mathrm{F}a_\mathrm{2D})$ vary from the centre to the outsides of the cloud. We relate these local quantities to those of a homogeneous system by applying the local density approximation (LDA). This approach is justified as long as the external potential varies weakly over the correlation length of the corresponding homogeneous system. In the normal phase this translates into the condition $k_\mathrm{F}(r)R\gg1$, where $R$ is the typical spatial extent of the cloud which in turn accounts for the scale of variation of the potential. For our typical experimental parameters, we can reliably apply the LDA up to around $T/T_F\approx 1.5$. Consequently we restrict our analysis of the pairing gap in Fig. 4 of the main text to this regime.

\begin{table*}[ht!]
	\centering
	\begin{tabular}{c|c|c|c|c|c|c}
		\parbox[c]{1.7cm}{Magnetic\\field (G)}& Mixture &\parbox[c]{1.7cm}{central\\$\ln(k_F a_{2D})$} &\parbox[c]{1.3cm}{central\\$T/T_F$}&\parbox{2.1cm}{two-body\\$E_B/h$ (kHz)}& \parbox[c]{2.1cm}{central\\$E_F/h$ (kHz)} &\parbox[c]{1.7cm}{central\\$\ln(k_F a_{2D})$\\ final state} \\
		\hline
		\hline
		672&$\ket{1}\ket{3}$&-0.16&0.50&9.31&7.61&-9.54\\
		682&$\ket{1}\ket{3}$&0.57&0.45&3.62&7.89 &-8.15\\
		687&$\ket{1}\ket{3}$&0.88&0.39&2.20&7.95&-7.53\\
		692&$\ket{1}\ket{3}$&1.06&0.44&1.37&6.56&-7.05\\
		822&$\ket{1}\ket{2}$&0.96&0.27&2.74&12.12&4.71\\
		834&$\ket{1}\ket{2}$&1.12&0.46&1.52&8.37&4.66\\
		854&$\ket{1}\ket{2}$&1.81&0.14&0.60&11.70&5.02\\
	\end{tabular}
	\caption{Experimental parameters for the datasets used in the main text.}
	\label{tab:T1_parameters}
\end{table*}

\subsection{Systematic Effects}
\textbf{Absorption imaging:} To determine the density $n_\mathrm{2D}\left(r\right)$ we use high intensity absorption imaging at intensities approximately equal to the saturation intensity. To compensate for the Doppler shift the atoms experience during the imaging, we perform a linear ramp of the laser frequency as described in \cite{Ries2015}. As the imaging transitions for state $\ket{1}$ and state $\ket{2}$ are not fully closed, especially at lower magnetic fields, we correct our obtained densities for dark state losses. Overall, our imaging calibration results in an uncertainty of $\sim\unit[7]{\%}$ in the density determination.\\

\noindent
\textbf{Atoms in adjacent layers of optical trap:} From tomographic RF measurements \cite{Ries2015} we can estimate that a small fraction  $ \lesssim \unit[10]{\%}$ of atoms is transferred into adjacent layers of the SWT. This results in an overestimation of the density and thus of our local Fermi momentum and temperature. We expect small additional contributions to the RF signal in the central region at intermediate frequencies due to the weakly populated layers.

\subsection{Experimental Parameters}
We quantify the different regimes of the BEC-BCS crossover by the local parameters $T/T_F$ and $\ln(k_Fa_{2D})$. The temperature $T$ of the sample is determined by the fitting procedure described above. The Fermi momentum and temperature are directly related to the measured local density via $k_F=\sqrt{4\pi n}$ and $T_F=\frac{\hbar^2 k_F^2}{2 m k_B}$, where $n$ is the atomic density of a single component. In addition, the 2D scattering length is obtained from the harmonic oscillator length $l_z=\sqrt{\hbar/m\omega_z}$ in z-direction and the 3D scattering length $a_{3D}$ following (\textit{7}) as
\begin{equation}
a_{2D}=l_z\sqrt{\frac{\pi}{A}}\exp\left(-\sqrt{\frac{\pi}{2}}\frac{l_z}{a_{3D}}\right),
\end{equation}
where $A\approx0.905$. We set the scattering length by adjusting the magnetic offset field in the vicinity of the broad Feshbach resonance at 690 G (832 G) in the $\ket{1}\ket{3}$ ($\ket{1}\ket{2}$) mixture as illustrated in Fig.\,S \ref{fig:S2_RFSpectroscopy} C. The magnetic field is calibrated by comparing the experimentally obtained free-free transition energies $E_\mathrm{free-free}$ in a spin-polarized and thus non-interacting sample to theoretical predictions using the Breit-Rabi formula.\

For the interpretation of our data, we compare the experimentally obtained pairing gap to the two-body bound state energy $E_\mathrm{B}$. It is calculated from the transcendental equation \cite{Levinsen2015}
\begin{equation}
\frac{l_z}{a}=\int_{0}^{\infty}\frac{du}{\sqrt{4\pi u^3}}\left(1-\frac{\exp(\frac{-E_B}{\hbar\omega_z}u)}{\sqrt{\frac{1}{2u}(1-\exp(-2u))}}\right).
\end{equation}
An overview over the experimental parameters can be found in Tab. S 1.\\

\section{Supplementary Information}

\subsection{Theoretical Background on Pairing in the BEC and BCS Limits}
In this section, we provide a brief account of the theoretical aspects of pairing in the well-understood BEC and BCS limits. The treatment presented here has been described in detail in the references \cite{Barth2014,Fischer2014}.\\
The BCS quasiparticle dispersion for an attractive Fermi gas is $E_k = \sqrt{\Delta^2+\xi_k^2}$, where $\Delta$ denotes the superconducting gap, and $\xi_k=E_k-\mu$ is the free dispersion relation $E_k=\hbar^2k^2/2m$ measured from the chemical potential $\mu$.  The spectral function has the form

\begin{equation}
\label{eq:ABCS}
A_\text{BCS}(k,E)
= \underbrace{v_k^2 \delta(E+E_k-\mu)}_\text{bound} + \underbrace{u_k^2 \delta(E-E_k-\mu)}_\text{free}
\end{equation}
for the bound and the free branches, with coherence factors $v_k^2 = (1-\xi_k/E_k)/2$ and $u_k^2 = (1+\xi_k/E_k)/2$.  Within 2D BCS theory the chemical potential at zero temperature $\mu=\EF - \EB/2$
and the gap $\Delta = \sqrt{2\EF\EB} = 2\EF e^{-\ln(\kfa)}$, while $T_c = e^\gamma\Delta/\pi = 1.13\,\EF\, e^{-\ln(\kfa)}$.\\

\noindent
On the BCS side $\EB\ll\EF$, $\mu>0$ and the dispersion reaches a
minimum gap $\Delta=E_{k_0}$ at wavevector $k_0 = \sqrt{2m\mu}/\hbar \approx k_F$, cf.\ Fig.~2C.  In the BEC limit $\EB\gg\EF$ one has $\mu\to-\EB/2$, and the coherence factor $v_k^2\to Z_k=k_F^2a_\text{2D}^2/(1+k^2a_\text{2D}^2)^2$ approaches the square of the bound-state wave function; thus the BCS spectral function asymptotically crosses over into BEC spectral function \cite{Barth2014}

\begin{equation}
\label{eq:ABEC}
A_\text{BEC}(k,E)
= \underbrace{Z_k \delta(E+\xi_k-\mu)}_\text{bound} + \underbrace{(1-Z_k)\delta(E-\xi_k-\mu)}_\text{free},
\end{equation}\\
\noindent
for the bound and the free branch.  The two branches bend outwards in Fig.\ref{fig:imbalance} B, with a minimal distance $\EB$ at $k=0$. \\

\noindent
The RF response in the absence of final-state interactions involving species $\ket c$ takes the form \cite{Fischer2014}

\begin{align}
\label{eq:rf1}
\Gamma_\text{BCS}(\omega)
= \pi\Omega^2 \sum_k A_\text{BCS}(k,E_k-\omega) f(\xi_k-\omega) \nonumber \\
= \frac{\pi\Omega^2N_b}{2\EF}\, f(\xi(\omega)-\omega)
\frac{\Delta^2}{\omega^2} \Theta(\xi(\omega)+\mu),
\end{align}
\\
\noindent
where $\Omega$ is the Rabi frequency, $f(E)$ the Fermi function, and $\xi(\omega)=(\omega^2-\Delta^2)/2\omega$.  $N_b$ denotes the number of particles in state $\ket b$, and we have used the constant density of states $\rho(E) = m/(2\pi\hbar^2)$ in 2D.  The $\Theta$ step function constrains the response to two branches: first, bound-free transitions from the lower branch occur only above a threshold frequency $\omega > \omega_\text{th} = \sqrt{\mu^2+\Delta^2}-\mu = \EB > 0$. Second, free-free transitions from the thermally excited upper branch appear at $\omega<0$, but are too weak to be observed at low temperature, cf.\ Fig.~1D.  Instead, for small spin imbalance a region around the minimum of the upper quasiparticle branch becomes occupied, and a new RF peak appears at $E_\text{free} = -\Delta$ \cite{Schirotzek2008}, cf.\ Fig.~ \ref{fig:imbalance} C,E.  Furthermore, if the interacting species $\ket b$ experiences a mean-field shift $U<0$, the
overall spectrum shifts to $E_\text{th} = \EB-U$ and $E_\text{free} = -\Delta-U$.  At zero temperature we thus obtain 

\begin{align}
\label{eq:13}
\Gamma_\text{BCS}(\omega)
= \frac{\pi\Omega^2N_b}{2\EF}\, \frac{\Delta^2}{\omega^2}
[ \Theta(\omega-\EB+U) \nonumber \\
 + \delta_\text{free}(\omega+\Delta+U) ].
\end{align}
\\
\noindent
{It is a remarkable consequence of pairing in 2D that, within the BCS theory, the RF threshold remains $\mathrm{E_\mathrm{th} = {\EB-U}}$ throughout the crossover. Hence, one cannot distinguish between two-body and many-body pairing by measuring the bound-free transition alone.}  Only by occupying the upper branch and measuring the free peak at $E_\text{free}=-\Delta-U$ can one determine the pairing energy $\Delta E=E_\text{th} - E_\text{free} = \EB+\Delta$ in a way that requires no knowledge of the mean-field shift $U$ and quantifies the many-body gap $\Delta$.

\subsection{Onset of Pairing}

\begin{figure*}[t!]
	\centering
	\includegraphics[width=0.6\textwidth]{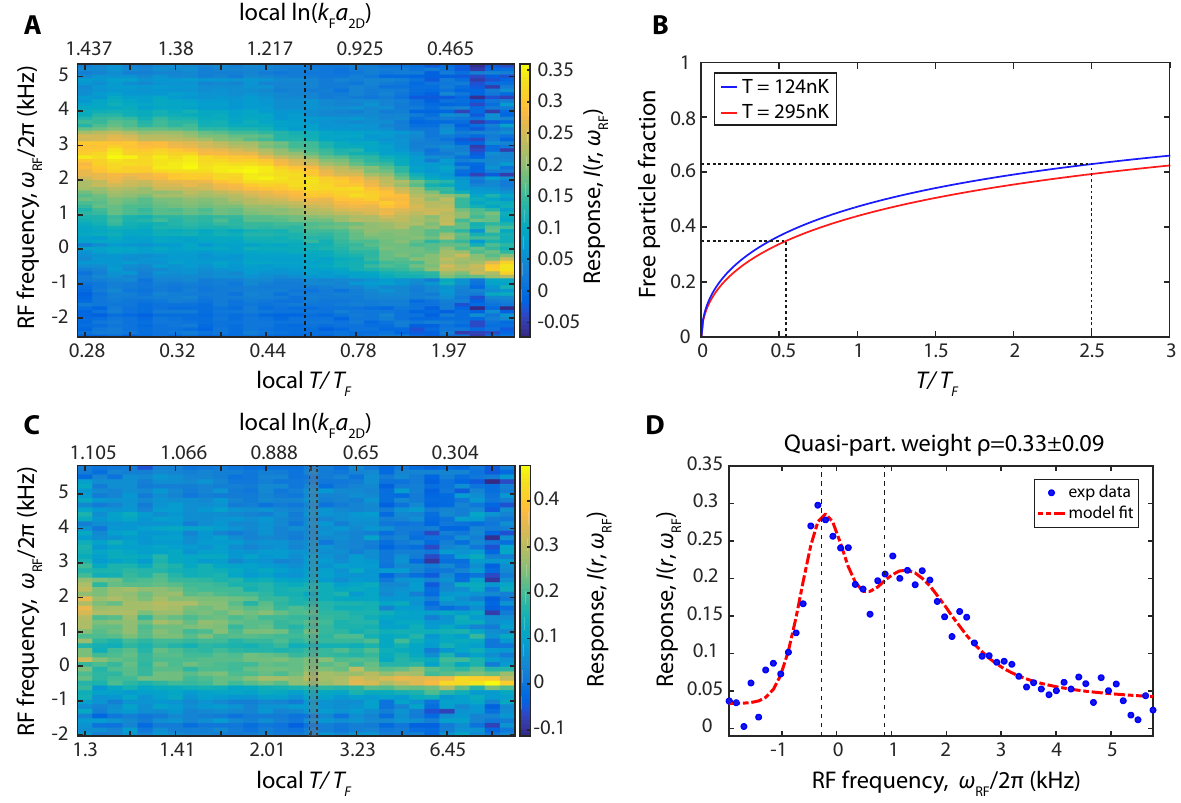}
	\caption{The onset of pairing in a balanced sample. \textbf{A}  and \textbf{C} show, for different temperatures regimes (T=\unit[124]{nK} and T=\unit[295]{nK}), the RF spectra in the $\ket{1}\ket{2}$-mixture at \unit[844]{G} plotted against RF frequency offset and radius, where the relative temperature and interaction parameter are indicated for a selection of radii.
		The vertical line in \textbf{A} marks the prediction for \unit[35]{\%} free particle fraction according to the Saha formula. The red curve in \textbf{B} represents the $\TTF$-dependence of the free-particle fraction for the parameters used in \textbf{A}. We see no significant free fraction at this radius. \textbf{D} shows a cut along the dotted line in \textbf{C} corresponding to $T/T_\mathrm{F} \approx 2.5$ where we observe a free particle fraction of $\sim\unit[35]{\%}$. Here, the Saha formula (blue curve in \textbf{B}) predicts more than \unit[60]{\%} unpaired particles.}
	\label{fig:S4_Weights}
\end{figure*}

The measurement of the pairing gap presented in the main text relies crucially on a sufficiently large population in both the paired and the quasi-particle branch. At finite temperature, quasiparticles are excited either thermally or by introducing a small imbalance. For the studied temperature range of $T/T_F \lesssim 1.5$ we do not observe a significant thermal occupation and thus we work with slight imbalance. This finding is illustrated in Fig.\,S \ref{fig:S4_Weights} A where the RF response of a spin-balanced sample is shown for an experimental setting corresponding to a central $\ln(k_Fa_{2D})\approx 1.4$. Indeed, there is no pronounced quasiparticle peak visible for $T/T_F \lesssim 1.5$. Nevertheless, as the phase-space-density decreases towards the wings of the cloud, a regime of mostly unpaired particles is eventually reached.

\noindent
For a non-interacting gas consisting of dimers and free atoms in thermal equilibrium, the unpaired fraction is given by the Saha formula \cite{Barth2014}:
\begin{equation}
\frac{n^2_f}{n_d}=\frac{mk_BT}{4\pi \hbar^2}e^{-\frac{E_B}{k_B T}},
\end{equation}
where $n_f$ ($n_d$) denotes the free (bound) density of particles. Indicated in Fig.\,S \ref{fig:S4_Weights} A is the relative temperature $T/T_F \approx 0.54$, where the Saha formula predicts a significant free particle fraction of \unit[35]{\%}. We observe an almost fully paired system at this temperature, however. To further study the onset of pairing in a density regime LDA is still valid, we heat the system by trap depth modulation. The corresponding RF spectrum is shown in Fig.\,S \ref{fig:S4_Weights} C. We observe a free particle fraction of $\sim \unit[35]{\%}$ only at around $T/T_F\approx2.5$.
While a quantitative analysis of the onset of pairing remains subject to further studies, the qualitative finding of significantly increased paired fractions compared to expectations from two-body theory further strengthens our interpretation of the normal phase above $T_c$ being strongly influenced by many-body correlations.

\subsection{Spectra in the Weakly Interacting Limit}

\begin{figure}[b!]
	\centering
	\includegraphics[width=0.4\textwidth]{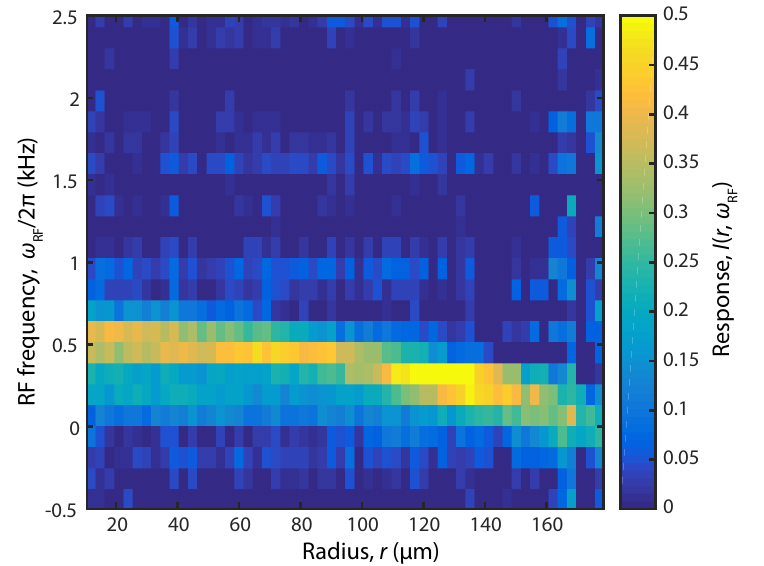}
	\caption{Spatially resolved spectra at $\ln{k_F a_{2D}} \sim 3.5$. The Fermi energy $E_\textrm{F}/h \approx 7\,$kHz at the center of the cloud. The fact that we see a single branch suggests the absence of many-body pairing in this regime.}
	\label{fig:S5_Highfield}
\end{figure}

In this section, we show the spatially resolved response function $I(r, \omega_\textrm{RF})$ measured at 1000 G with a $\ket{1}\ket{2}$-mixture, which corresponds to $\ln(k_\textrm{F} a_\textrm{2D}) \sim 3.5$ and therefore represents the weakly interacting limit of the 2D BEC-BCS crossover. At this interaction strength, the binding energy is $\EB/h \approx 7\,$Hz, which is much smaller than all  thermodynamic scales and our experimental RF resolution. In  Fig.\,S \ref{fig:S5_Highfield}, we observe only a single branch in the spectrum. This branch appears close to $ \omega_\textrm{RF} \sim 0$ and has a maximum value of approximately 500 Hz at the center where the Fermi energy ($\EF/h$) is about $\unit[7]{kHz}$. The response also exhibits a density-dependent mean-field shift as expected from an interacting system of fermions. \\
Although it is difficult to make a reliable statement about pairing in this regime due to our finite resolution, the presence of a single branch does suggest the absence of a gap that is larger than the scale of our resolution, or in other words $\Delta E \lesssim 0.05\,E_\textrm{F}$ at the center of the trap. This should be contrasted with the spectra at $\ln(k_\textrm{F} a_\textrm{2D}) \sim 1$, where the measured $\Delta E \approx 0.6\,E_\textrm{F}$.


\begin{thebibliography}{32}%
	\makeatletter
	\providecommand \@ifxundefined [1]{%
		\@ifx{#1\undefined}
	}%
	\providecommand \@ifnum [1]{%
		\ifnum #1\expandafter \@firstoftwo
		\else \expandafter \@secondoftwo
		\fi
	}%
	\providecommand \@ifx [1]{%
		\ifx #1\expandafter \@firstoftwo
		\else \expandafter \@secondoftwo
		\fi
	}%
	\providecommand \natexlab [1]{#1}%
	\providecommand \enquote  [1]{``#1''}%
	\providecommand \bibnamefont  [1]{#1}%
	\providecommand \bibfnamefont [1]{#1}%
	\providecommand \citenamefont [1]{#1}%
	\providecommand \href@noop [0]{\@secondoftwo}%
	\providecommand \href [0]{\begingroup \@sanitize@url \@href}%
	\providecommand \@href[1]{\@@startlink{#1}\@@href}%
	\providecommand \@@href[1]{\endgroup#1\@@endlink}%
	\providecommand \@sanitize@url [0]{\catcode `\\12\catcode `\$12\catcode
		`\&12\catcode `\#12\catcode `\^12\catcode `\_12\catcode `\%12\relax}%
	\providecommand \@@startlink[1]{}%
	\providecommand \@@endlink[0]{}%
	\providecommand \url  [0]{\begingroup\@sanitize@url \@url }%
	\providecommand \@url [1]{\endgroup\@href {#1}{\urlprefix }}%
	\providecommand \urlprefix  [0]{URL }%
	\providecommand \Eprint [0]{\href }%
	\providecommand \doibase [0]{http://dx.doi.org/}%
	\providecommand \selectlanguage [0]{\@gobble}%
	\providecommand \bibinfo  [0]{\@secondoftwo}%
	\providecommand \bibfield  [0]{\@secondoftwo}%
	\providecommand \translation [1]{[#1]}%
	\providecommand \BibitemOpen [0]{}%
	\providecommand \bibitemStop [0]{}%
	\providecommand \bibitemNoStop [0]{.\EOS\space}%
	\providecommand \EOS [0]{\spacefactor3000\relax}%
	\providecommand \BibitemShut  [1]{\csname bibitem#1\endcsname}%
	\let\auto@bib@innerbib\@empty
	\bibitem [{\citenamefont {Yang}(1962)}]{Yang1962}%
	\BibitemOpen
	\bibfield  {author} {\bibinfo {author} {\bibfnamefont {C.~N.}\ \bibnamefont
			{Yang}},\ }\href {\doibase 10.1103/RevModPhys.34.694} {\bibfield  {journal}
		{\bibinfo  {journal} {Rev. Mod. Phys.}\ }\textbf {\bibinfo {volume} {34}},\
		\bibinfo {pages} {694} (\bibinfo {year} {1962})}\BibitemShut {NoStop}%
	\bibitem [{\citenamefont {Bardeen}\ \emph {et~al.}(1957)\citenamefont
		{Bardeen}, \citenamefont {Cooper},\ and\ \citenamefont
		{Schrieffer}}]{Bardeen1957}%
	\BibitemOpen
	\bibfield  {author} {\bibinfo {author} {\bibfnamefont {J.}~\bibnamefont
			{Bardeen}}, \bibinfo {author} {\bibfnamefont {L.~N.}\ \bibnamefont {Cooper}},
		\ and\ \bibinfo {author} {\bibfnamefont {J.~R.}\ \bibnamefont {Schrieffer}},\
	}\href {\doibase 10.1103/PhysRev.108.1175} {\bibfield  {journal} {\bibinfo
			{journal} {Phys. Rev.}\ }\textbf {\bibinfo {volume} {108}},\ \bibinfo {pages}
		{1175} (\bibinfo {year} {1957})}\BibitemShut {NoStop}%
	\bibitem [{\citenamefont {Chen}\ \emph {et~al.}(2009)\citenamefont {Chen},
		\citenamefont {He}, \citenamefont {Chien},\ and\ \citenamefont
		{Levin}}]{Chen2009}%
	\BibitemOpen
	\bibfield  {author} {\bibinfo {author} {\bibfnamefont {Q.}~\bibnamefont
			{Chen}}, \bibinfo {author} {\bibfnamefont {Y.}~\bibnamefont {He}}, \bibinfo
		{author} {\bibfnamefont {C.-C.}\ \bibnamefont {Chien}}, \ and\ \bibinfo
		{author} {\bibfnamefont {K.}~\bibnamefont {Levin}},\ }\href {\doibase
		10.1088/0034-4885/72/12/122501} {\bibfield  {journal} {\bibinfo  {journal}
			{Reports on Progress in Physics}\ }\textbf {\bibinfo {volume} {72}},\
		\bibinfo {pages} {122501} (\bibinfo {year} {2009})}\BibitemShut {NoStop}%
	\bibitem [{\citenamefont {Mueller}()}]{Mueller2017}%
	\BibitemOpen
	\bibfield  {author} {\bibinfo {author} {\bibfnamefont {E.~J.}\ \bibnamefont
			{Mueller}},\ }\href@noop {} {\ }\Eprint
	{http://arxiv.org/abs/arXiv:1701.04838v1} {arXiv:1701.04838v1} \BibitemShut
	{NoStop}%
	\bibitem [{\citenamefont {Lee}\ \emph {et~al.}(2014)\citenamefont {Lee},
		\citenamefont {Schmitt}, \citenamefont {Moore}, \citenamefont {Johnston},
		\citenamefont {Cui}, \citenamefont {Li}, \citenamefont {Yi}, \citenamefont
		{Liu}, \citenamefont {Hashimoto}, \citenamefont {Zhang}, \citenamefont {Lu},
		\citenamefont {Devereaux}, \citenamefont {Lee},\ and\ \citenamefont
		{Shen}}]{Lee2014}%
	\BibitemOpen
	\bibfield  {author} {\bibinfo {author} {\bibfnamefont {J.~J.}\ \bibnamefont
			{Lee}}, \bibinfo {author} {\bibfnamefont {F.~T.}\ \bibnamefont {Schmitt}},
		\bibinfo {author} {\bibfnamefont {R.~G.}\ \bibnamefont {Moore}}, \bibinfo
		{author} {\bibfnamefont {S.}~\bibnamefont {Johnston}}, \bibinfo {author}
		{\bibfnamefont {Y.-T.}\ \bibnamefont {Cui}}, \bibinfo {author} {\bibfnamefont
			{W.}~\bibnamefont {Li}}, \bibinfo {author} {\bibfnamefont {M.}~\bibnamefont
			{Yi}}, \bibinfo {author} {\bibfnamefont {Z.~K.}\ \bibnamefont {Liu}},
		\bibinfo {author} {\bibfnamefont {M.}~\bibnamefont {Hashimoto}}, \bibinfo
		{author} {\bibfnamefont {Y.}~\bibnamefont {Zhang}}, \bibinfo {author}
		{\bibfnamefont {D.~H.}\ \bibnamefont {Lu}}, \bibinfo {author} {\bibfnamefont
			{T.~P.}\ \bibnamefont {Devereaux}}, \bibinfo {author} {\bibfnamefont {D.-H.}\
			\bibnamefont {Lee}}, \ and\ \bibinfo {author} {\bibfnamefont {Z.-X.}\
			\bibnamefont {Shen}},\ }\href {\doibase doi:10.1038/nature13894} {\bibfield
		{journal} {\bibinfo  {journal} {Nature}\ }\textbf {\bibinfo {volume} {515}},\
		\bibinfo {pages} {245} (\bibinfo {year} {2014})}\BibitemShut {NoStop}%
	\bibitem [{\citenamefont {Kasahara}\ \emph {et~al.}(2016)\citenamefont
		{Kasahara}, \citenamefont {Yamashita}, \citenamefont {Shi}, \citenamefont
		{Kobayashi}, \citenamefont {Shimoyama}, \citenamefont {Watashige},
		\citenamefont {Ishida}, \citenamefont {Terashima}, \citenamefont {Wolf},
		\citenamefont {Hardy}, \citenamefont {Meingast}, \citenamefont
		{v.~Löhneysen}, \citenamefont {Levchenko}, \citenamefont {Shibauchi},\ and\
		\citenamefont {Matsuda}}]{Kasahara2016}%
	\BibitemOpen
	\bibfield  {author} {\bibinfo {author} {\bibfnamefont {S.}~\bibnamefont
			{Kasahara}}, \bibinfo {author} {\bibfnamefont {T.}~\bibnamefont {Yamashita}},
		\bibinfo {author} {\bibfnamefont {A.}~\bibnamefont {Shi}}, \bibinfo {author}
		{\bibfnamefont {R.}~\bibnamefont {Kobayashi}}, \bibinfo {author}
		{\bibfnamefont {Y.}~\bibnamefont {Shimoyama}}, \bibinfo {author}
		{\bibfnamefont {T.}~\bibnamefont {Watashige}}, \bibinfo {author}
		{\bibfnamefont {K.}~\bibnamefont {Ishida}}, \bibinfo {author} {\bibfnamefont
			{T.}~\bibnamefont {Terashima}}, \bibinfo {author} {\bibfnamefont
			{T.}~\bibnamefont {Wolf}}, \bibinfo {author} {\bibfnamefont {F.}~\bibnamefont
			{Hardy}}, \bibinfo {author} {\bibfnamefont {C.}~\bibnamefont {Meingast}},
		\bibinfo {author} {\bibfnamefont {H.}~\bibnamefont {v.~Löhneysen}}, \bibinfo
		{author} {\bibfnamefont {A.}~\bibnamefont {Levchenko}}, \bibinfo {author}
		{\bibfnamefont {T.}~\bibnamefont {Shibauchi}}, \ and\ \bibinfo {author}
		{\bibfnamefont {Y.}~\bibnamefont {Matsuda}},\ }\href {\doibase
		doi:10.1038/ncomms12843} {\bibfield  {journal} {\bibinfo  {journal} {Nature
				Communications}\ }\textbf {\bibinfo {volume} {7}},\ \bibinfo {pages} {12843}
		(\bibinfo {year} {2016})}\BibitemShut {NoStop}%
	\bibitem [{\citenamefont {Levinsen}\ and\ \citenamefont
		{Parish}(2015)}]{Levinsen2015}%
	\BibitemOpen
	\bibfield  {author} {\bibinfo {author} {\bibfnamefont {J.}~\bibnamefont
			{Levinsen}}\ and\ \bibinfo {author} {\bibfnamefont {M.~M.}\ \bibnamefont
			{Parish}},\ }\href
	{http://www.worldscientific.com/doi/abs/10.1142/9789814667746_0001}
	{\bibfield  {journal} {\bibinfo  {journal} {Annual Review of Cold Atoms and
				Molecules}\ }\textbf {\bibinfo {volume} {3}},\ \bibinfo {pages} {1} (\bibinfo
		{year} {2015})}\BibitemShut {NoStop}%
	\bibitem [{\citenamefont {Ries}\ \emph {et~al.}(2015)\citenamefont {Ries},
		\citenamefont {Wenz}, \citenamefont {Z\"urn}, \citenamefont {Bayha},
		\citenamefont {Boettcher}, \citenamefont {Kedar}, \citenamefont {Murthy},
		\citenamefont {Neidig}, \citenamefont {Lompe},\ and\ \citenamefont
		{Jochim}}]{Ries2015}%
	\BibitemOpen
	\bibfield  {author} {\bibinfo {author} {\bibfnamefont {M.~G.}\ \bibnamefont
			{Ries}}, \bibinfo {author} {\bibfnamefont {A.~N.}\ \bibnamefont {Wenz}},
		\bibinfo {author} {\bibfnamefont {G.}~\bibnamefont {Z\"urn}}, \bibinfo
		{author} {\bibfnamefont {L.}~\bibnamefont {Bayha}}, \bibinfo {author}
		{\bibfnamefont {I.}~\bibnamefont {Boettcher}}, \bibinfo {author}
		{\bibfnamefont {D.}~\bibnamefont {Kedar}}, \bibinfo {author} {\bibfnamefont
			{P.~A.}\ \bibnamefont {Murthy}}, \bibinfo {author} {\bibfnamefont
			{M.}~\bibnamefont {Neidig}}, \bibinfo {author} {\bibfnamefont
			{T.}~\bibnamefont {Lompe}}, \ and\ \bibinfo {author} {\bibfnamefont
			{S.}~\bibnamefont {Jochim}},\ }\href {\doibase
		10.1103/PhysRevLett.114.230401} {\bibfield  {journal} {\bibinfo  {journal}
			{Phys. Rev. Lett.}\ }\textbf {\bibinfo {volume} {114}},\ \bibinfo {pages}
		{230401} (\bibinfo {year} {2015})}\BibitemShut {NoStop}%
	\bibitem [{\citenamefont {Murthy}\ \emph {et~al.}(2015)\citenamefont {Murthy},
		\citenamefont {Boettcher}, \citenamefont {Bayha}, \citenamefont {Holzmann},
		\citenamefont {Kedar}, \citenamefont {Neidig}, \citenamefont {Ries},
		\citenamefont {Wenz}, \citenamefont {Z\"urn},\ and\ \citenamefont
		{Jochim}}]{Murthy2015}%
	\BibitemOpen
	\bibfield  {author} {\bibinfo {author} {\bibfnamefont {P.~A.}\ \bibnamefont
			{Murthy}}, \bibinfo {author} {\bibfnamefont {I.}~\bibnamefont {Boettcher}},
		\bibinfo {author} {\bibfnamefont {L.}~\bibnamefont {Bayha}}, \bibinfo
		{author} {\bibfnamefont {M.}~\bibnamefont {Holzmann}}, \bibinfo {author}
		{\bibfnamefont {D.}~\bibnamefont {Kedar}}, \bibinfo {author} {\bibfnamefont
			{M.}~\bibnamefont {Neidig}}, \bibinfo {author} {\bibfnamefont {M.~G.}\
			\bibnamefont {Ries}}, \bibinfo {author} {\bibfnamefont {A.~N.}\ \bibnamefont
			{Wenz}}, \bibinfo {author} {\bibfnamefont {G.}~\bibnamefont {Z\"urn}}, \ and\
		\bibinfo {author} {\bibfnamefont {S.}~\bibnamefont {Jochim}},\ }\href
	{\doibase 10.1103/PhysRevLett.115.010401} {\bibfield  {journal} {\bibinfo
			{journal} {Phys. Rev. Lett.}\ }\textbf {\bibinfo {volume} {115}},\ \bibinfo
		{pages} {010401} (\bibinfo {year} {2015})}\BibitemShut {NoStop}%
	\bibitem [{\citenamefont {T\"{o}rm\"{a}}(2016)}]{Torma2016}%
	\BibitemOpen
	\bibfield  {author} {\bibinfo {author} {\bibfnamefont {P.}~\bibnamefont
			{T\"{o}rm\"{a}}},\ }\href {\doibase 10.1088/0031-8949/91/4/043006} {\bibfield
		{journal} {\bibinfo  {journal} {Physica Scripta}\ }\textbf {\bibinfo
			{volume} {91}},\ \bibinfo {pages} {043006} (\bibinfo {year}
		{2016})}\BibitemShut {NoStop}%
	\bibitem [{\citenamefont {Nascimb\`ene}\ \emph {et~al.}(2011)\citenamefont
		{Nascimb\`ene}, \citenamefont {Navon}, \citenamefont {Pilati}, \citenamefont
		{Chevy}, \citenamefont {Giorgini}, \citenamefont {Georges},\ and\
		\citenamefont {Salomon}}]{Nascimbene2011}%
	\BibitemOpen
	\bibfield  {author} {\bibinfo {author} {\bibfnamefont {S.}~\bibnamefont
			{Nascimb\`ene}}, \bibinfo {author} {\bibfnamefont {N.}~\bibnamefont {Navon}},
		\bibinfo {author} {\bibfnamefont {S.}~\bibnamefont {Pilati}}, \bibinfo
		{author} {\bibfnamefont {F.}~\bibnamefont {Chevy}}, \bibinfo {author}
		{\bibfnamefont {S.}~\bibnamefont {Giorgini}}, \bibinfo {author}
		{\bibfnamefont {A.}~\bibnamefont {Georges}}, \ and\ \bibinfo {author}
		{\bibfnamefont {C.}~\bibnamefont {Salomon}},\ }\href {\doibase
		10.1103/PhysRevLett.106.215303} {\bibfield  {journal} {\bibinfo  {journal}
			{Phys. Rev. Lett.}\ }\textbf {\bibinfo {volume} {106}},\ \bibinfo {pages}
		{215303} (\bibinfo {year} {2011})}\BibitemShut {NoStop}%
	\bibitem [{\citenamefont {Gaebler}\ \emph {et~al.}(2010)\citenamefont
		{Gaebler}, \citenamefont {Stewart}, \citenamefont {Drake}, \citenamefont
		{Jin}, \citenamefont {Perali}, \citenamefont {Pieri},\ and\ \citenamefont
		{Strinati}}]{Gaebler2010}%
	\BibitemOpen
	\bibfield  {author} {\bibinfo {author} {\bibfnamefont {J.~P.}\ \bibnamefont
			{Gaebler}}, \bibinfo {author} {\bibfnamefont {J.~T.}\ \bibnamefont
			{Stewart}}, \bibinfo {author} {\bibfnamefont {T.~E.}\ \bibnamefont {Drake}},
		\bibinfo {author} {\bibfnamefont {D.~S.}\ \bibnamefont {Jin}}, \bibinfo
		{author} {\bibfnamefont {A.}~\bibnamefont {Perali}}, \bibinfo {author}
		{\bibfnamefont {P.}~\bibnamefont {Pieri}}, \ and\ \bibinfo {author}
		{\bibfnamefont {G.~C.}\ \bibnamefont {Strinati}},\ }\href {\doibase
		10.1038/NPHYS1709} {\bibfield  {journal} {\bibinfo  {journal} {Nature
				Physics}\ }\textbf {\bibinfo {volume} {6}},\ \bibinfo {pages} {569} (\bibinfo
		{year} {2010})}\BibitemShut {NoStop}%
	\bibitem [{\citenamefont {Perali}\ \emph {et~al.}(2011)\citenamefont {Perali},
		\citenamefont {Palestini}, \citenamefont {Pieri}, \citenamefont {Strinati},
		\citenamefont {Stewart}, \citenamefont {Gaebler}, \citenamefont {Drake},\
		and\ \citenamefont {Jin}}]{Perali2011}%
	\BibitemOpen
	\bibfield  {author} {\bibinfo {author} {\bibfnamefont {A.}~\bibnamefont
			{Perali}}, \bibinfo {author} {\bibfnamefont {F.}~\bibnamefont {Palestini}},
		\bibinfo {author} {\bibfnamefont {P.}~\bibnamefont {Pieri}}, \bibinfo
		{author} {\bibfnamefont {G.~C.}\ \bibnamefont {Strinati}}, \bibinfo {author}
		{\bibfnamefont {J.~T.}\ \bibnamefont {Stewart}}, \bibinfo {author}
		{\bibfnamefont {J.~P.}\ \bibnamefont {Gaebler}}, \bibinfo {author}
		{\bibfnamefont {T.~E.}\ \bibnamefont {Drake}}, \ and\ \bibinfo {author}
		{\bibfnamefont {D.~S.}\ \bibnamefont {Jin}},\ }\href {\doibase
		10.1103/PhysRevLett.106.060402} {\bibfield  {journal} {\bibinfo  {journal}
			{Phys. Rev. Lett.}\ }\textbf {\bibinfo {volume} {106}},\ \bibinfo {pages}
		{060402} (\bibinfo {year} {2011})}\BibitemShut {NoStop}%
	\bibitem [{\citenamefont {Sagi}\ \emph {et~al.}(2015)\citenamefont {Sagi},
		\citenamefont {Drake}, \citenamefont {Paudel}, \citenamefont {Chapurin},\
		and\ \citenamefont {Jin}}]{Sagi2015}%
	\BibitemOpen
	\bibfield  {author} {\bibinfo {author} {\bibfnamefont {Y.}~\bibnamefont
			{Sagi}}, \bibinfo {author} {\bibfnamefont {T.~E.}\ \bibnamefont {Drake}},
		\bibinfo {author} {\bibfnamefont {R.}~\bibnamefont {Paudel}}, \bibinfo
		{author} {\bibfnamefont {R.}~\bibnamefont {Chapurin}}, \ and\ \bibinfo
		{author} {\bibfnamefont {D.~S.}\ \bibnamefont {Jin}},\ }\href {\doibase
		10.1103/PhysRevLett.114.075301} {\bibfield  {journal} {\bibinfo  {journal}
			{Phys. Rev. Lett.}\ }\textbf {\bibinfo {volume} {114}},\ \bibinfo {pages}
		{075301} (\bibinfo {year} {2015})}\BibitemShut {NoStop}%
	\bibitem [{\citenamefont {Schunck}\ \emph {et~al.}(2007)\citenamefont
		{Schunck}, \citenamefont {Shin}, \citenamefont {Schirotzek}, \citenamefont
		{Zwierlein},\ and\ \citenamefont {Ketterle}}]{Schunck2007}%
	\BibitemOpen
	\bibfield  {author} {\bibinfo {author} {\bibfnamefont {C.~H.}\ \bibnamefont
			{Schunck}}, \bibinfo {author} {\bibfnamefont {Y.}~\bibnamefont {Shin}},
		\bibinfo {author} {\bibfnamefont {A.}~\bibnamefont {Schirotzek}}, \bibinfo
		{author} {\bibfnamefont {M.~W.}\ \bibnamefont {Zwierlein}}, \ and\ \bibinfo
		{author} {\bibfnamefont {W.}~\bibnamefont {Ketterle}},\ }\href {\doibase
		10.1126/science.1140749} {\bibfield  {journal} {\bibinfo  {journal}
			{Science}\ }\textbf {\bibinfo {volume} {316}},\ \bibinfo {pages} {867}
		(\bibinfo {year} {2007})}\BibitemShut {NoStop}%
	\bibitem [{\citenamefont {Sommer}\ \emph {et~al.}(2012)\citenamefont {Sommer},
		\citenamefont {Cheuk}, \citenamefont {Ku}, \citenamefont {Bakr},\ and\
		\citenamefont {Zwierlein}}]{Sommer2012}%
	\BibitemOpen
	\bibfield  {author} {\bibinfo {author} {\bibfnamefont {A.~T.}\ \bibnamefont
			{Sommer}}, \bibinfo {author} {\bibfnamefont {L.~W.}\ \bibnamefont {Cheuk}},
		\bibinfo {author} {\bibfnamefont {M.~J.~H.}\ \bibnamefont {Ku}}, \bibinfo
		{author} {\bibfnamefont {W.~S.}\ \bibnamefont {Bakr}}, \ and\ \bibinfo
		{author} {\bibfnamefont {M.~W.}\ \bibnamefont {Zwierlein}},\ }\href {\doibase
		10.1103/PhysRevLett.108.045302} {\bibfield  {journal} {\bibinfo  {journal}
			{Phys. Rev. Lett.}\ }\textbf {\bibinfo {volume} {108}},\ \bibinfo {pages}
		{045302} (\bibinfo {year} {2012})}\BibitemShut {NoStop}%
	\bibitem [{\citenamefont {Feld}\ \emph {et~al.}(2011)\citenamefont {Feld},
		\citenamefont {Fr\"{o}hlich}, \citenamefont {Vogt}, \citenamefont
		{Koschorreck},\ and\ \citenamefont {K\"{o}hl}}]{Feld2011}%
	\BibitemOpen
	\bibfield  {author} {\bibinfo {author} {\bibfnamefont {M.}~\bibnamefont
			{Feld}}, \bibinfo {author} {\bibfnamefont {B.}~\bibnamefont {Fr\"{o}hlich}},
		\bibinfo {author} {\bibfnamefont {E.}~\bibnamefont {Vogt}}, \bibinfo {author}
		{\bibfnamefont {M.}~\bibnamefont {Koschorreck}}, \ and\ \bibinfo {author}
		{\bibfnamefont {M.}~\bibnamefont {K\"{o}hl}},\ }\href {\doibase
		10.1038/nature10627} {\bibfield  {journal} {\bibinfo  {journal} {Nature}\
		}\textbf {\bibinfo {volume} {480}},\ \bibinfo {pages} {75} (\bibinfo {year}
		{2011})}\BibitemShut {NoStop}%
	\bibitem [{\citenamefont {Z\"urn}\ \emph {et~al.}(2013)\citenamefont {Z\"urn},
		\citenamefont {Lompe}, \citenamefont {Wenz}, \citenamefont {Jochim},
		\citenamefont {Julienne},\ and\ \citenamefont {Hutson}}]{Zurn2013}%
	\BibitemOpen
	\bibfield  {author} {\bibinfo {author} {\bibfnamefont {G.}~\bibnamefont
			{Z\"urn}}, \bibinfo {author} {\bibfnamefont {T.}~\bibnamefont {Lompe}},
		\bibinfo {author} {\bibfnamefont {A.~N.}\ \bibnamefont {Wenz}}, \bibinfo
		{author} {\bibfnamefont {S.}~\bibnamefont {Jochim}}, \bibinfo {author}
		{\bibfnamefont {P.~S.}\ \bibnamefont {Julienne}}, \ and\ \bibinfo {author}
		{\bibfnamefont {J.~M.}\ \bibnamefont {Hutson}},\ }\href {\doibase
		10.1103/PhysRevLett.110.135301} {\bibfield  {journal} {\bibinfo  {journal}
			{Phys. Rev. Lett.}\ }\textbf {\bibinfo {volume} {110}},\ \bibinfo {pages}
		{135301} (\bibinfo {year} {2013})}\BibitemShut {NoStop}%
	\bibitem [{SOM()}]{SOM2017}%
	\BibitemOpen
	\href@noop {} {\bibinfo  {journal} {Supplementary Materials}\ }\BibitemShut
	{NoStop}%
	\bibitem [{EFn()}]{EFnote2017}%
	\BibitemOpen
	\bibfield  {journal} {  }\href@noop {} {\bibinfo  {journal} {$E_\text{F}/h$ has
			a typical value of 7.5 kHz at the center of the cloud}\ }\BibitemShut
	{NoStop}%
	\bibitem [{\citenamefont {Boettcher}\ \emph {et~al.}(2016)\citenamefont
		{Boettcher}, \citenamefont {Bayha}, \citenamefont {Kedar}, \citenamefont
		{Murthy}, \citenamefont {Neidig}, \citenamefont {Ries}, \citenamefont {Wenz},
		\citenamefont {Z\"urn}, \citenamefont {Jochim},\ and\ \citenamefont
		{Enss}}]{Boettcher2016}%
	\BibitemOpen
	\bibfield  {journal} {  }\bibfield  {author} {\bibinfo {author} {\bibfnamefont
			{I.}~\bibnamefont {Boettcher}}, \bibinfo {author} {\bibfnamefont
			{L.}~\bibnamefont {Bayha}}, \bibinfo {author} {\bibfnamefont
			{D.}~\bibnamefont {Kedar}}, \bibinfo {author} {\bibfnamefont {P.~A.}\
			\bibnamefont {Murthy}}, \bibinfo {author} {\bibfnamefont {M.}~\bibnamefont
			{Neidig}}, \bibinfo {author} {\bibfnamefont {M.~G.}\ \bibnamefont {Ries}},
		\bibinfo {author} {\bibfnamefont {A.~N.}\ \bibnamefont {Wenz}}, \bibinfo
		{author} {\bibfnamefont {G.}~\bibnamefont {Z\"urn}}, \bibinfo {author}
		{\bibfnamefont {S.}~\bibnamefont {Jochim}}, \ and\ \bibinfo {author}
		{\bibfnamefont {T.}~\bibnamefont {Enss}},\ }\href {\doibase
		10.1103/PhysRevLett.116.045303} {\bibfield  {journal} {\bibinfo  {journal}
			{Phys. Rev. Lett.}\ }\textbf {\bibinfo {volume} {116}},\ \bibinfo {pages}
		{045303} (\bibinfo {year} {2016})}\BibitemShut {NoStop}%
	\bibitem [{\citenamefont {Shin}\ \emph {et~al.}(2007)\citenamefont {Shin},
		\citenamefont {Schunck}, \citenamefont {Schirotzek},\ and\ \citenamefont
		{Ketterle}}]{Shin2007}%
	\BibitemOpen
	\bibfield  {author} {\bibinfo {author} {\bibfnamefont {Y.}~\bibnamefont
			{Shin}}, \bibinfo {author} {\bibfnamefont {C.~H.}\ \bibnamefont {Schunck}},
		\bibinfo {author} {\bibfnamefont {A.}~\bibnamefont {Schirotzek}}, \ and\
		\bibinfo {author} {\bibfnamefont {W.}~\bibnamefont {Ketterle}},\ }\href
	{\doibase 10.1103/PhysRevLett.99.090403} {\bibfield  {journal} {\bibinfo
			{journal} {Phys. Rev. Lett.}\ }\textbf {\bibinfo {volume} {99}},\ \bibinfo
		{pages} {090403} (\bibinfo {year} {2007})}\BibitemShut {NoStop}%
	\bibitem [{\citenamefont {Schirotzek}\ \emph {et~al.}(2008)\citenamefont
		{Schirotzek}, \citenamefont {Shin}, \citenamefont {Schunck},\ and\
		\citenamefont {Ketterle}}]{Schirotzek2008}%
	\BibitemOpen
	\bibfield  {author} {\bibinfo {author} {\bibfnamefont {A.}~\bibnamefont
			{Schirotzek}}, \bibinfo {author} {\bibfnamefont {Y.-i.}\ \bibnamefont
			{Shin}}, \bibinfo {author} {\bibfnamefont {C.~H.}\ \bibnamefont {Schunck}}, \
		and\ \bibinfo {author} {\bibfnamefont {W.}~\bibnamefont {Ketterle}},\ }\href
	{\doibase 10.1103/PhysRevLett.101.140403} {\bibfield  {journal} {\bibinfo
			{journal} {Phys. Rev. Lett.}\ }\textbf {\bibinfo {volume} {101}},\ \bibinfo
		{pages} {140403} (\bibinfo {year} {2008})}\BibitemShut {NoStop}%
	\bibitem [{\citenamefont {Barth}\ and\ \citenamefont
		{Hofmann}(2014)}]{Barth2014}%
	\BibitemOpen
	\bibfield  {author} {\bibinfo {author} {\bibfnamefont {M.}~\bibnamefont
			{Barth}}\ and\ \bibinfo {author} {\bibfnamefont {J.}~\bibnamefont
			{Hofmann}},\ }\href {\doibase 10.1103/PhysRevA.89.013614} {\bibfield
		{journal} {\bibinfo  {journal} {Phys. Rev. A}\ }\textbf {\bibinfo {volume}
			{89}},\ \bibinfo {pages} {013614} (\bibinfo {year} {2014})}\BibitemShut
	{NoStop}%
	\bibitem [{\citenamefont {Fischer}\ and\ \citenamefont
		{Parish}(2014)}]{Fischer2014}%
	\BibitemOpen
	\bibfield  {author} {\bibinfo {author} {\bibfnamefont {A.~M.}\ \bibnamefont
			{Fischer}}\ and\ \bibinfo {author} {\bibfnamefont {M.~M.}\ \bibnamefont
			{Parish}},\ }\href {\doibase 10.1103/PhysRevB.90.214503} {\bibfield
		{journal} {\bibinfo  {journal} {Phys. Rev. B}\ }\textbf {\bibinfo {volume}
			{90}},\ \bibinfo {pages} {214503} (\bibinfo {year} {2014})}\BibitemShut
	{NoStop}%
	\bibitem [{\citenamefont {Bauer}\ \emph {et~al.}(2014)\citenamefont {Bauer},
		\citenamefont {Parish},\ and\ \citenamefont {Enss}}]{Bauer2014}%
	\BibitemOpen
	\bibfield  {author} {\bibinfo {author} {\bibfnamefont {M.}~\bibnamefont
			{Bauer}}, \bibinfo {author} {\bibfnamefont {M.~M.}\ \bibnamefont {Parish}}, \
		and\ \bibinfo {author} {\bibfnamefont {T.}~\bibnamefont {Enss}},\ }\href
	{\doibase 10.1103/PhysRevLett.112.135302} {\bibfield  {journal} {\bibinfo
			{journal} {Phys. Rev. Lett.}\ }\textbf {\bibinfo {volume} {112}},\ \bibinfo
		{pages} {135302} (\bibinfo {year} {2014})}\BibitemShut {NoStop}%
	\bibitem [{\citenamefont {Vitali}\ \emph {et~al.}()\citenamefont {Vitali},
		\citenamefont {Shi}, \citenamefont {Qin},\ and\ \citenamefont
		{Zhang}}]{Vitali2017}%
	\BibitemOpen
	\bibfield  {author} {\bibinfo {author} {\bibfnamefont {E.}~\bibnamefont
			{Vitali}}, \bibinfo {author} {\bibfnamefont {H.}~\bibnamefont {Shi}},
		\bibinfo {author} {\bibfnamefont {M.}~\bibnamefont {Qin}}, \ and\ \bibinfo
		{author} {\bibfnamefont {S.}~\bibnamefont {Zhang}},\ }\href@noop {} {\
	}\Eprint {http://arxiv.org/abs/arXiv:1705.07929v1} {arXiv:1705.07929v1}
	\BibitemShut {NoStop}%
	\bibitem [{\citenamefont {Langmack}\ \emph {et~al.}(2012)\citenamefont
		{Langmack}, \citenamefont {Barth}, \citenamefont {Zwerger},\ and\
		\citenamefont {Braaten}}]{Langmack2012}%
	\BibitemOpen
	\bibfield  {author} {\bibinfo {author} {\bibfnamefont {C.}~\bibnamefont
			{Langmack}}, \bibinfo {author} {\bibfnamefont {M.}~\bibnamefont {Barth}},
		\bibinfo {author} {\bibfnamefont {W.}~\bibnamefont {Zwerger}}, \ and\
		\bibinfo {author} {\bibfnamefont {E.}~\bibnamefont {Braaten}},\ }\href
	{\doibase 10.1103/PhysRevLett.108.060402} {\bibfield  {journal} {\bibinfo
			{journal} {Phys. Rev. Lett.}\ }\textbf {\bibinfo {volume} {108}},\ \bibinfo
		{pages} {060402} (\bibinfo {year} {2012})}\BibitemShut {NoStop}%
	\bibitem [{\citenamefont {Fr\"ohlich}\ \emph {et~al.}(2012)\citenamefont
		{Fr\"ohlich}, \citenamefont {Feld}, \citenamefont {Vogt}, \citenamefont
		{Koschorreck}, \citenamefont {K\"{o}hl}, \citenamefont {Berthod},\ and\
		\citenamefont {Giamarchi}}]{Frohlich2012}%
	\BibitemOpen
	\bibfield  {author} {\bibinfo {author} {\bibfnamefont {B.}~\bibnamefont
			{Fr\"ohlich}}, \bibinfo {author} {\bibfnamefont {M.}~\bibnamefont {Feld}},
		\bibinfo {author} {\bibfnamefont {E.}~\bibnamefont {Vogt}}, \bibinfo {author}
		{\bibfnamefont {M.}~\bibnamefont {Koschorreck}}, \bibinfo {author}
		{\bibfnamefont {M.}~\bibnamefont {K\"{o}hl}}, \bibinfo {author}
		{\bibfnamefont {C.}~\bibnamefont {Berthod}}, \ and\ \bibinfo {author}
		{\bibfnamefont {T.}~\bibnamefont {Giamarchi}},\ }\href {\doibase
		10.1103/PhysRevLett.109.130403} {\bibfield  {journal} {\bibinfo  {journal}
			{Phys. Rev. Lett.}\ }\textbf {\bibinfo {volume} {109}},\ \bibinfo {pages}
		{130403} (\bibinfo {year} {2012})}\BibitemShut {NoStop}%
	\bibitem [{\citenamefont {Ngampruetikorn}\ \emph {et~al.}(2013)\citenamefont
		{Ngampruetikorn}, \citenamefont {Levinsen},\ and\ \citenamefont
		{Parish}}]{Ngampruetikorn2013}%
	\BibitemOpen
	\bibfield  {author} {\bibinfo {author} {\bibfnamefont {V.}~\bibnamefont
			{Ngampruetikorn}}, \bibinfo {author} {\bibfnamefont {J.}~\bibnamefont
			{Levinsen}}, \ and\ \bibinfo {author} {\bibfnamefont {M.~M.}\ \bibnamefont
			{Parish}},\ }\href {\doibase 10.1103/PhysRevLett.111.265301} {\bibfield
		{journal} {\bibinfo  {journal} {Phys. Rev. Lett.}\ }\textbf {\bibinfo
			{volume} {111}},\ \bibinfo {pages} {265301} (\bibinfo {year}
		{2013})}\BibitemShut {NoStop}%
	\bibitem [{\citenamefont {Mitra}\ \emph {et~al.}(2016)\citenamefont {Mitra},
		\citenamefont {Brown}, \citenamefont {Schau\ss{}}, \citenamefont {Kondov},\
		and\ \citenamefont {Bakr}}]{Mitra2016}%
	\BibitemOpen
	\bibfield  {author} {\bibinfo {author} {\bibfnamefont {D.}~\bibnamefont
			{Mitra}}, \bibinfo {author} {\bibfnamefont {P.~T.}\ \bibnamefont {Brown}},
		\bibinfo {author} {\bibfnamefont {P.}~\bibnamefont {Schau\ss{}}}, \bibinfo
		{author} {\bibfnamefont {S.~S.}\ \bibnamefont {Kondov}}, \ and\ \bibinfo
		{author} {\bibfnamefont {W.~S.}\ \bibnamefont {Bakr}},\ }\href {\doibase
		10.1103/PhysRevLett.117.093601} {\bibfield  {journal} {\bibinfo  {journal}
			{Phys. Rev. Lett.}\ }\textbf {\bibinfo {volume} {117}},\ \bibinfo {pages}
		{093601} (\bibinfo {year} {2016})}\BibitemShut {NoStop}%
	\bibitem [{\citenamefont {Koschorreck}\ \emph {et~al.}(2012)\citenamefont
		{Koschorreck}, \citenamefont {Pertot}, \citenamefont {Vogt}, \citenamefont
		{Fr\"{o}hlich}, \citenamefont {Feld},\ and\ \citenamefont
		{K\"{o}hl}}]{Koschorreck2012}%
	\BibitemOpen
	\bibfield  {author} {\bibinfo {author} {\bibfnamefont {M.}~\bibnamefont
			{Koschorreck}}, \bibinfo {author} {\bibfnamefont {D.}~\bibnamefont {Pertot}},
		\bibinfo {author} {\bibfnamefont {E.}~\bibnamefont {Vogt}}, \bibinfo {author}
		{\bibfnamefont {B.}~\bibnamefont {Fr\"{o}hlich}}, \bibinfo {author}
		{\bibfnamefont {M.}~\bibnamefont {Feld}}, \ and\ \bibinfo {author}
		{\bibfnamefont {M.}~\bibnamefont {K\"{o}hl}},\ }\href {\doibase
		10.1038/nature11151} {\bibfield  {journal} {\bibinfo  {journal} {Nature}\
		}\textbf {\bibinfo {volume} {485}},\ \bibinfo {pages} {619} (\bibinfo {year}
		{2012})}\BibitemShut {NoStop}%
\end{thebibliography}
\end{document}